\documentclass[superscriptaddress,groupedaddress,nofootnoteinbib,11pt]{article}  

\pdfoutput=1
\usepackage[pdftex]{graphics}
\usepackage{epsfig,amssymb,bm,graphicx}
\usepackage{amsmath}
\usepackage[usenames,dvipsnames]{color}
	\usepackage{cancel}

\usepackage{slashed}

\usepackage{graphicx}  
\usepackage{dcolumn}   
\usepackage{bm}        
\usepackage{slashed}        
\usepackage{float}
\usepackage{jcappub}
\usepackage{color}

\usepackage{graphicx}
\usepackage{dcolumn}
\usepackage{bm}
\usepackage{amssymb}
\usepackage{amsmath}
\usepackage{sectsty}
\usepackage{colortbl}
\usepackage{latexsym}
\usepackage{float}
\usepackage{ifthen}
\usepackage{enumerate}
\usepackage{url}
\usepackage[force]{feynmp-auto}
\usepackage{jcappub}
    \usepackage{picinpar}
    \usepackage{colortbl}
\usepackage{multirow}
	\usepackage{float}
	      \usepackage{setspace}
\usepackage{array}
\usepackage{bm}
\usepackage{amsopn}

\usepackage{caption}
\usepackage{subcaption}

\usepackage{booktabs}
\usepackage[table]{xcolor}

\begin{document}

\title{Primordial Black Hole Production in Natural and Hilltop Inflation}

\author[a]{Jessica L. Cook}

\affiliation[a]{Department of Physical Sciences, Eastern Connecticut State University, \\ 83 Windham St, Willimantic, USA}

\emailAdd{cookje@easternct.edu}
\date{\today}
\abstract{

We consider the possibility of primordial black hole, PBH, formation sourced by a rise in the power spectrum. The power spectrum becomes large at late times due to decay of the inflaton into vectors through a $\phi F \tilde{F}$ coupling. Two background inflaton models which are well supported by current Planck data are considered, natural inflation and hilltop inflation. Many of the papers considering formation of PBHs have considered a peaked power spectrum where $P_{\zeta}$ gets small again at late times. This avoids overproducing miniature PBHs which would evaporate and could violate BBN and CMB bounds. This paper examines the other way of avoiding these bounds, producing PBHs from perturbations formed closer to the end of inflation such that the PBHs are too small to violate these bounds. This has the advantage of allowing for simpler models in that no additional feature is needed to be added to evade constraints. Although these black holes would have evaporated, they can be close to without exceeding current BBN bounds, making it possible the signature will be revealed in the future. We calculate how the various model parameters affect the mass and number of PBHs produced. Any evidence for PBHs sourced from an inflationary power spectrum would provide evidence for inflation on a drastically different energy scale from the CMB, and thus would be highly valuable in answering what occurred during inflation.

 }
\maketitle

\section{Introduction}

Since LIGO first started observing black hole collisions, the community has wondered if all the observed black holes are astrophysical (produced from stellar colalpse), or if at least some of them are primordial (forming from overdensities generated in the early universe) \cite{PhysRevLett.116.201301}. There are various reasons why LIGO's black hole distribution seemed curious. In the simplest stellar collapse in a binary system, one expects the black hole partners to be of similar mass to each other, in the range of 20-50 solar masses, and have aligned spins \cite{LIGOScientific:2020kqk}. While the majority of LIGO's events are within that range, they have seen events above this limit. In particular, if astrophysical, neither binary partner is expected in  the pair-instability gap starting at around 50 solar masses \cite{Abbott:2020tfl,LIGOScientific:2020kqk}. Stars with cores above this limit are expected to leave no remenant. \cite{LIGOScientific:2020kqk} lists three events where one black hole has a mass above 45 solar masses, one of which has a mass around 85 solar masses. However, it's been suggested that uncertainties in the calculation could mean the actual lower mass limit might be as high as 70 solar masses \cite{Woosley:2021xba}.

Early LIGO events showed spin statistics different from what's expected for astrophyscial black holes. The simpliest stellar collapsed binary partners are expected to have ailged spins. However, early LIGO events fit best with a BH distribution largely of BHs with low spin and isotropically aligned spins and orbits \cite{Fernandez:2019kyb}. Small magnitude and isotropically aligned spins are what is predicted from primordial black holes, PBHs. This isn't conclusive as there are various reasons why astrophysical BHs might not have the simplest expected spin statistics. For example, BHs that form inside dense clusters are expected to show a larger difference in partner masses and spin alignments.

There are multiple reasons why finding PBHs would be especially significant. Any unevaporated PBHs would automatically constitute at least some of the dark matter. It's been shown that PBHs would likely be only a small component of the dark matter, however, there are narrow mass bands between $10^{17}- 10^{23}$ g and $10 -100$ solar masses where PBHs could still make up the majority of dark matter \cite{Carr:2021bzv}.  Assuming the density fluctuations which created the PBHs were generated during inflation, observing PBHs would provide a window into inflation at much higher frequencies than is observed from CMB fluctuations. One of the strongest limits on our ability to confirm inflation or answer how it happened is the limited observational evidence from this time. Extra data on inflation at a very different scale might finally answer these questions. 

Thus there are a lot of reasons why it's interesting to ask if PBHs exist. This paper explores the possibility of creating PBHs from peaks in the inflationary scalar power spectrum. These peaks would become regions of overdensity and when of sufficient size, would collapse into PBHs during reheating or the subsequent radiation dominated period.  To generate observable PBHs, one typically needs a power spectrum around $10^{-3}$. However, the power spectrum at CMB scales is only around $2 \times 10^{-9}$. One needs to explain why the inflationary power spectrum would grow dramatically after CMB scales exited the horizon. More common inflationary models have the power spectrum gradually decreasing  due to a diminishing Hubble parameter and a gradually increasing rolling of the inflaton field during inflation leading to a gradually increasing slow roll parameter $\epsilon$.  

Something non-standard has to happen to create this rapidly growing power spectrum. That typically isn't sufficient though. Bounds on PBHs are also strong at scales corresponding to high frequencies/ late times during inflation. The higher the frequency scale when a black hole forms, the lower the mass of the black hole. Sufficiently small black holes will have evaporated due to Hawking radiation, and this radiation can have significant effects on the predictions of big bang nucleousythesis, BBN, the CMB, CMB anisotripoies, and the intergalacic and galactic gamma ray background \cite{Carr:2020gox}.  This leads to some of the strongest bounds on PBHs.  This evaporation cutoff typically occurs around 15 efolds before the end of inflation, although it varies a little from model to model.  This means PBHs are both strongly constrained at low frequencies corresponding to where CMB scales are observable and at high frequencies where Hawking radiation constraints are relevant. For this reason, many papers seek to generate a peaked power spectrum to evade these constraints.

This paper will consider generating PBH production from inflation through a $\phi F \tilde{F}$ coupling where a pseudoscalar inflaton decays into vector fields. This causes an increase in the power spectrum at high frequencies leading to possible PBH formation. Such a model is also interesting for other reasons. The model also produces significant equilateral non-Gaussianities and a unique chiral and unusually large/ maybe detectable gravitational wave signal \cite{Anber:2009ua}. A variety of other papers have considered PBH formation from such an intereation. Some of these papers have placed constraints on the power spectrum based on the nonobservation of PBHs \cite{Domcke:2016bkh, Erfani:2015rqv} often assuming a $m^2 \phi^2$ potential for $\phi$ \cite{Linde:2012bt, Garcia-Bellido:2016dkw, Bugaev:2013fya} or a Starobinsky potential \cite{Garcia-Bellido:2016dkw} or a linear potential \cite{Bugaev:2013fya}. \cite{Lin:2012gs} used natural inflation but didn't include backreaction. 

Various groups have looked at modifying the inflaton potential in some way, usually to increase $\dot{\phi}$ at intermediate scales followed by a decrease at higher frequency scales, to either get observational gravitational waves at direct detection experiments and/or large PBHs without overproducing PBHs at small scales. See for example \cite{Garcia-Bellido:2016dkw, Cheng:2016qzb, Garcia-Bellido:2017mdw, Motohashi:2017kbs, Ballesteros:2017fsr}. These complications could be adding additional scalar fields \cite{Garcia-Bellido:2016dkw} or adding extra terms to the potential to change its shape where needed as in inflection point models \cite{Cheng:2016qzb, Ozsoy:2020kat, Cheng:2018yyr} or by adding a non-minimal gravitational coupling to the inflaton \cite{Domcke:2016bkh, Domcke:2017fix}. It could be adding non-cannonical kinetic terms as in $k$ inflation \cite{Lin:2020goi, Solbi:2021wbo, Lin:2021vwc}. \cite{Cheng:2015oqa, Cheng:2018yyr, Domcke:2020zez} have done a careful treatment of the $\phi F \tilde{F}$ coupling in the backreaction regime finding oscillations in the inflaton’s velocity and as a consequence in $P_{\zeta}$.

One needs to consider a specific background inflaton model in order to make concrete predictions. The standard, simple model people used to turn to was $m^2 \phi^2$, but this appears ruled out from non-observation of CMB B modes. Limiting the tensor to scalar ratio $r$ to the current $2\sigma$ bound of $0.036$ from Planck, WMAP BICEP, and Keck would correspond to $n_s=0.991$ which is more than $6 \sigma$ removed from the observed $n_s$ \cite{BICEP:2021xfz}.  In this paper, I'll consider two other models which fit the current Planck data well: natural inflation and quartic hilltop inflation. Both models can produce Planck's observed scalar spectral index $n_s$ while also producing a small yet to be observed tensor to scalar ratio $r$, and while staying consistent with reasonable assumptions about reheating.

Most groups have focused on trying to get intermediate mass PBHs produced from features in the power spectrum roughly in the middle of inflation. This allows one to avoid the strong constraints at CMB scales. This is also an interesting region as depending on the model, a peak in the power spectrum could coincide with a peak in the tensor spectrum and so could coincide with observable gravitational waves in a gravitational wave detector like LIGO or LISA. The difficulty is, one often then has to add something to the model to make the power spectrum drop after this phenomenologically interesting region. This is needed to avoid the strong constraints at high frequencies arising from the effect of small, evaporating PBHs on BBN and the CMB. However, this typically makes the models more complicated. There is another possibility. The peak in the power spectrum could come later, towards the end of inflation and coincide with and modify BBN predictions. This largely unexplored region of parameter space is what this paper will focus on. 

There are two elements measured to high precision which agree with the theoretical BBN predictions, deuterium, D, and $^4$He. The primordial $^3$He abundance is hard to measure because it's both produced and destroyed in stars. While this is also true of $^4$He, there are methods for estimating $^4$He back to the primordial abundance. One can also measure the lithium abundance, but here the prediction and measured quantity disagree. Less Li is observed than expected. It's possible an extended mass spectrum of PBHs might ease this tension \cite{Carr:2020gox}. 

Small mass PBHs, $m < 10^9$ g, evaporate before weak force freeze out have no effect on BBN and are unconstrained. However, evaporation of PBHs with mass around $10^9  < m < 10^{13}$ g evaporate during or just after BBN will affect the primordial abundances. The models below will produce PBHs right in this mass range which evaporate during BBN. The various particles produced from this evaporation and the secondary particles (produced from decay of particles produced from evaporation) will alter the primordial abundances in different ways. Evaporation can have multiple competing effects, so predicting the final abundances from the power spectrum is complicated. One typically runs a computer simulation  with all relevant cross sections to account for all the possible effects \cite{Carr:2020gox, Auffinger:2022khh}. However, certain interactions are more significant than others for different PBHs mass. 


Two background inflaton models are considered below, natural inflation and hilltop inflation. The natural inflation scenario will face the largest constraints from PBH masses in the range $10^{10.5} - 10^{13.1}$ g. In this range, the most significant effect of evaporation is from hadrons and photons breaking apart $^4$He, which ultimately results in extra D and $^3$He. For hilltop inflation, the strongest bounds will come from evaporating PBHs of mass $10^9$ to $10^{9.8}$ g. Here the most significant effect will be from prolonging conversion of protons to neutrons, past when weak scale freeze out should have occurred, leading to extra neutrons and ultimately extra $^4$He \cite{Carr:2020gox}. 


While there is agreement between the predictions and measured quantities for $^4$He and D, they are continually improving in precision. A disagreement could be uncovered in the future indicated new physics. The theoretical predictions for the abundances are dependent on knowledge of a variety of cross sections, and the precision of these cross sections is gradually improving. For example, \cite{Yeh:2020mgl} shows an improvement in the predicted deuterium to hydrogen abundance, D/H $= 2.51 \pm 0.11 \times 10^{-5}$, after using an improved $^3$He cross section measured by the LUNA collaboration.





The experimental deuterium abundance is measured as a fraction of the hydrogen abundance using spectroscopy on high redshift gas clouds illuminated by high redshift quasars \cite{Riemer-Sorensen:2017pey}. Both are typically around redshift 3.5. These measurements have been done multiple times using different quasar gas cloud pairs. The resultant average D measurement is gradually improving in precision as the cumulative observation time on these systems increases. \cite{Yeh:2020mgl} gives D/H $= 2.55 \pm 0.03 \times 10^-5$ as an average over 11 separate observations.  

The primordial $^4$He abundance, $Y_P$, is measured from the emission spectrum of hot gas clouds in low metallicity star forming HII (ionized hydrogen) regions and galaxies. The abundance of materials produced in supernovas are also measured (O, N, or S), and the He abundance is extrapolated back to a no metals, primordial abundance. Results are gradually improving. For example, groups are tracing the abundance of more elements like N and S where originally only O was used \cite{Fern_ndez_2018}. Further, data modeling is continually improving as groups add extra spectral lines to their fits \cite{Aver:2020fon}. \cite{Aver:2020fon} report a primordial helium mass fraction of $Y_P = 0.2453 \pm 0.0034$. As both the theory and experimental values for D and $^4$He improve, it's possible a discrepancy will be observed implying new physics. 

\section{ Energy Density in PBHs: }

A common approximation for the mass of PBHs which form at a particular time is given by the total mass encased in the universe at the time times a proportionality factor $\gamma$. $\gamma$ should be around $.2- .4$. For example see \cite{Inomata:2016rbd, Garcia-Bellido:2017mdw}.  I'll use $\gamma = 0.2$.

\begin{align}\label{eq5}
M_{PBH} \Big|_{formation} = \gamma \cdot\rho \cdot \frac{4}{3} \pi R_p^3 
\end{align}

\noindent $R_p$ is the physical horizon size, $\frac{1}{H}$, and $\rho$ is the mass density.  Note since $H$ is always decreasing, the sooner PBHs are produced, the smaller they will be. For perspective, I'll be interested in PBHs forming during the radiation dominated period after inflation. 

\begin{figure} 
\centering
    \includegraphics[width=8cm]{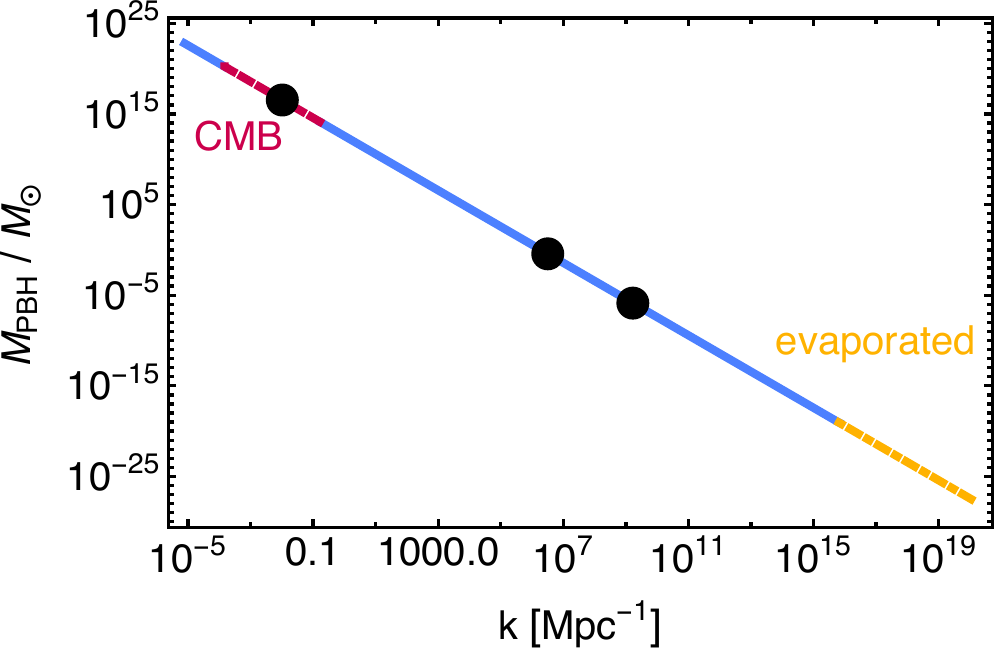}
\caption{This shows the relationship of eq.~(\ref{eq5}). The red, dotted segment on the left shows scales corresponding to the CMB, $l = 2$ to $l = 2500$. The orange, dash dotted segment on the right represents black holes which formed so early, they would be small enough to have evaporated before today. The black dots from left to right mark the scales of matter-radiation equality, QCD symmetry breaking, and electroweak symmetry breaking (assumed to occur at 100 GeV).}
\label{fig:a}
\end{figure}

The relationship of eq. \ref{eq5} is shown in Fig.~\ref{fig:a} where the $x$ axis plots the comoving frequency $k=\frac{a}{a_0 R_p}$. The red region shows CMB scales from $l = 2$ to $l = 2500$. The yellow shows PBHs that would have evaporated before today. The three black dots from small $k$ to large depict the scales of matter radiation equality, QCD symmetry breaking, and electroweak symmetry breaking. The later two are included because various groups have suggested models where PBHs coulud be produced connected with those events. Matter radiation equality is indicated to reinforce the assumption that the PBHs formed during the radiation dominated period.  If a perturbation generated during inflation later sourced a PBH right at matter radiation equality, it would have formed in the middle of when the CMB scales were freezing out of the horizon.  PBHs which formed later in the matter dominated period would correlate to perturbations  formed at lower frequencies to the left of that dot. Since we don't see peaks in the power spectrum on CMB scales, one doesn't expect to see PBHs sourced from this period, at least if inflation is the cause. Perturbations generated during inflation on scales to the right of this dot would lead to PBHs produced earlier after the end of inflation during the radiation dominated period or even earlier during reheating.

It's convenient to rewrite eq. \ref{eq5} in terms of the comoving horizon size $R = R_p/a$ and a variety of known constants. First we can use the first Friedmann equation: $H^2 = \frac{\rho}{3 M_P^2}$. If one assumes that the black holes form during the radiation dominated period, then the total energy density is related to the temperature by:

\begin{align}
\rho = \frac{\pi^2 g}{30} T^4 
\end{align}

\noindent where $g$ is the number of relativistic degrees of freedom. Assuming entropy is conserved from black hole formation to matter radiation equality and that  relativistic particles carry the vast majority of the entropy gives:

\begin{align}
a \,g^{1/3}\, T = a_{eq}\, g_{eq}^{1/3}\, T_{eq}
\end{align}

\noindent All non-subscript variables should be understood to be evaluated at black hole formation and the subscript $eq$ means evaluated at matter radiation equality. Finally I'll use

\begin{align}
\rho_{eq} = 2 \frac{\pi^2 g_{eq}}{30} T_{eq}^4 
\end{align}

\noindent where the factor of two accounts for the fact that relativistic particles make up half the energy density at equality. This gives:

\begin{align}
M_{PBH} =  2 \sqrt{\frac{2}{3}}  \pi \gamma M_P  \left(\frac{g_{eq}}{g} \right)^{1/6} a_{eq}^2 \sqrt{\rho_{eq}} R^2
 \end{align}
 
\noindent $g_{eq} = 3.36$  when the only relativisitic degrees of freedom left were photons and neutrinos. Throughout the paper, I'll use Planck's 2018 results using TT, TE, EE, lowE, lensing, and BAO: $\frac{a_0}{a_{eq}} = 3388$ and $\rho_{eq} = 8.96 \times 10^{-37}$ GeV$^4$  \cite{Aghanim:2018eyx}.  The following gives $M_{PBH}$ in GeV assuming $R$ is in $\frac{1}{GeV}$:

\begin{align}\label{eq8}
M_{PBH} =  1.26 \times 10^{-6}  \, \frac{\gamma }{g^{1/6}} \,  (a_0  R)^2
 \end{align}

These are black holes which formed long after inflation ended. However, assuming these PBHs formed from perturbations sourced during inflation, it's useful to rewrite eq. \ref{eq8} in terms of efollding time, $N$. This will be the efolding time during inflation when these perturbations first froze out of the horizon. I'll use that at horizon crossing $R = \frac{1}{aH}$. Let the subscript $i$ represent quantities evaluated during inflation when the perturbation in question formed. Let the subscript $f$ represent quantities evaluated much later when the black hole formed. Note $R_i = R_f$; the black holes form when the perturbation which sources them is able to reenter the horizon.  I'll choose the efolding number $N$ to be positive and count down to 0 at the end of inflation. To keep the inflation formulas simple, I'll use $a_{end} = 1$ where end means the end of inflation so that $a$ at any point during inflation is simply given by $e^{-N}$.

   \begin{align}\label{eq3}
M_{PBH} = 1.26 \times 10^{-6}  \, \frac{\gamma }{g^{1/6}} \,  \left(\frac{a_0}{a_{end} \,H_i}\, e^{N} \right)^2
 \end{align}

 Using $a_{end}=1$ means I cannot use the convention $a_0$, meaning $a$ today, $ = 1$ and must solve for $a_0$. This depends on how long reheating lasts which depends on the inflation model being used.  Using that $\frac{a_1}{a_2} = e^{\Delta N}$,

\begin{align}\label{eq9}
\frac{a_{end}}{a_0} = \frac{a_{eq}}{a_0} e^{- N_{re}} e^{- N_{RD}}
\end{align}

\noindent where $a_{eq}$ is evaluated at matter radiation equality, $N_{re}$ is the length of reheating, and $N_{RD}$ is the length of the subsequent radiation dominated epoch.  It looks like the result should be dependent on reheating assumptions. However, once an inflation model is chosen, the sum $N_{re} + N_{RD}$ is fixed. The sum $N_{re} + N_{RD}$ is given by (see for example \cite{Dai:2014jja, Cook:2015vqa}):

\begin{align}\label{eqw}
N_{re} + N_{RD} = \ln \left(\frac{a_{eq}}{a_0} \right) - N_k  - \ln \left(\frac{k_{pivot}}{a_0\, H_k} \right)
\end{align}

\noindent $H_k$ and $N_k$ are evaluated at a chosen pivot scale where matching can be made onto CMB data for $A_s$ and $n_s$. I'll use Planck's pivot of 0.05 Mpc$^{-1}$ such that $\frac{k_{pivot}}{a_0} = 0.05\, \frac{1}{\mbox{Mpc}}$.

\begin{figure}
\centering
\begin{subfigure}{.5\textwidth}
  \centering
  \includegraphics[width=.95\linewidth]{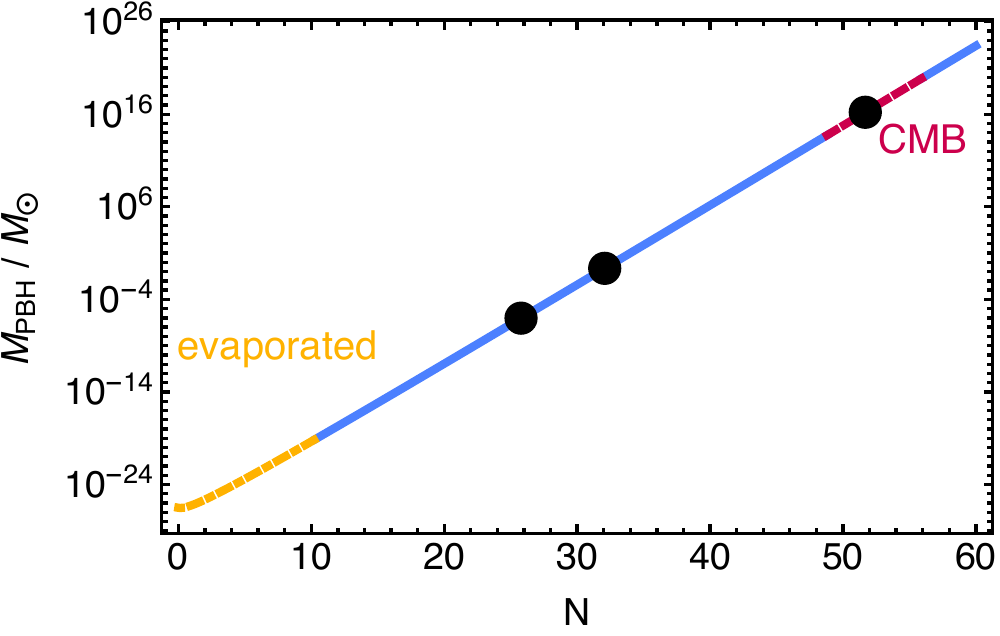}
\end{subfigure}%
\begin{subfigure}{.5\textwidth}
  \centering
  \includegraphics[width=.95\linewidth]{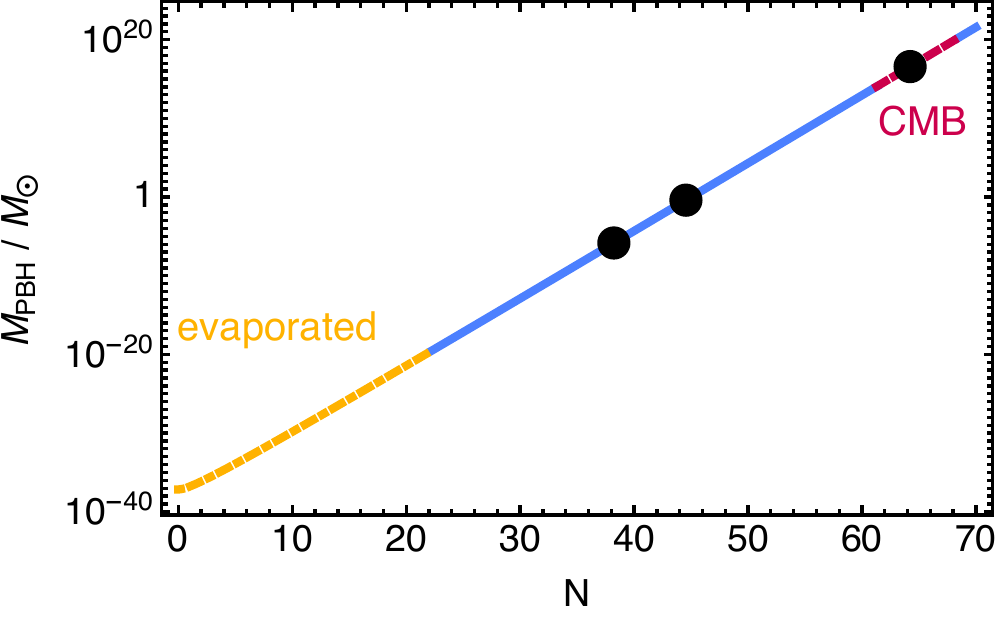}
\end{subfigure}
\caption{Each figure shows the correspondence between when a perturbation formed during inflation and the resulting PBH mass. The figure on the left shows the result for quartic hilltop inflation with $\mu = 30 \, M_P$ and $M = 1.55 \times 10^{16}$ GeV which will be considered below. The figure on the right shows the same for  natural inflation with $f = 9.45\, M_P$  and $\lambda = 1.77 \times 10^{16}$ GeV. Both have parameters and initial conditions chosen to give Planck's observed $n_s$ and $A_s$. Both show only the background $\phi$ model without a vector coupling. $N$ is defined to count down to 0 at the end of inflation. The regions and markings indicated are the same as in Fig.~\ref{fig:a}.}
\label{fig:b}
\end{figure}

Fig.~\ref{fig:a}   has no early universe model dependence. It is also interesting to make the same kind of figure but on the $x$-axis show $N$ using eq. \ref{eq3}. This is model dependent. However, one can get a feeling for when a certain black hole mass perturbation formed by showing the result for a couple popular inflation models. I'll use the two models which I'll focus on later in the paper, natural inflation and quartic hilltop inflation. Fig.~\ref{fig:b} shows this relationship for hilltop inflation on the left and natural inflation on the right. $N$ is defined to count down to 0 at the end of inflation.  Both models have parameters chosen to match Planck's observed $n_s$ and $A_s$ values. The markings are the same as in Fig.~\ref{fig:a} . The three black dots from large $N$ to small depict the scales of matter radiation equality, QCD symmetry breaking, electroweak symmetry breaking. The red region shows CMB scales from $l = 2$ to $l = 2500$. The yellow shows PBHs that would have evaporated by today.

Notice in the figure that $M_{PBH}$ is smaller, for a given $N$ value, for the natural inflation scenario compared to the hilltop scenario. This is because $M_{PBH}$ is proportional to $e^{N_{re} + N_{RD}}$ and $N_{re} + N_{RD} = 64.23$ in the hilltop scenario and only 52.15 in the natural scenario for the inflation parameters chosen here. The exponential conveys this into a large difference in $M_{PBH}$. The evaporation cutoff occurs at the same $k$ value for each, $8 \times 10^{-24}$ GeV, see the appendix, but this threshold occurs earlier in the natural inflation scenario at $N = 24$ vs. $N = 12$ in the hilltop scenario.

\subsection{Mass Fraction $\beta$}

We have an expression for $M_{PBH} (N)$. Next I'll write an expression for what's typically termed $\beta$, the fraction of the mass of the universe which collapsed into black holes. Once we have $\beta(N)$ we can combine it with $M_{PBH} (N)$ to get $\beta(M_{PBH})$.

First we need the probability distribution function of density perturbations. The $\phi F \tilde{F}$ coupling which through decay of $\phi$ into vectors is generating large peaks in the inflationary power spectrum also generates large equilateral non-Gaussianities. This has been shown by vaious authors; see for example \cite{Cook:2013xea}. The probability distribution of density perturbations of amplitude $\delta$ for such equilateral non-Gaussianities is given by:

\begin{align}
\mbox{PDF} = \frac{1}{\sqrt{2  \pi } \, \sigma\, \sqrt{\delta + \sigma^2}} e^{ - \frac{(\delta + \sigma^2)}{2 \,\sigma^2}}
\end{align}

\noindent where $\sigma$ is the variance; see for example \cite{Linde:2012bt, Bugaev:2013fya, Domcke:2017fix}. This gives the fraction of perturbations with amplitude $\delta$. Following the Press-Schechter formalism, we then count the fraction of perturbations larger than the critical density perturbation $\delta_c$ and call this $\beta$, \cite{Press:1973iz}. These can be assumed to have collapsed into black holes. 

\begin{align}
\beta = \int_{\delta_c}^{\infty} d \delta \, \, \mbox{PDF} 
\end{align}

\begin{align}
\beta_f (R) = \frac{1}{\sqrt{2 \pi } \, \sigma} \int_{\delta_c}^{\infty} d \delta \, \frac{1}{\sqrt{\delta + \sigma^2}} \, e^{ - \frac{(\delta + \sigma^2)}{2 \, \sigma^2}}
\end{align}

\noindent I'm writing a subscript $f$ to emphasize that $\beta$ is evaluated at the time of PBH formation, when the forming PBHs are the size of the horizon. Therefore, $\beta$ is a function of PBH size.  This can be simplified:

\begin{align}
\beta = \mbox{Erfc} \left( \sqrt{\frac{1}{2} \left(\frac{\delta_c}{\sigma^2} + 1 \right)  }  \right)
\end{align}

 The final mass prediction in PBHs is sensitive to $\delta_c$. It's not known exactly what $\delta_c$ should be, and it changes over time as the pressure of the primordial fluid changes. We'll take during the radiation dominated period $\delta_c = 0.41$ \cite{Harada:2013epa}.

One must next relate $\sigma$ to $P_{\zeta}$. $\sigma$ should technically be the mass variance, a function of the smoothed density power spectrum. However, one can instead approximate $\sigma$ as a function of the primordial, inflationary power spectrum.  It's been shown that this approximation produces results off by order 10\% \cite{Byrnes:2012yx}. 

Focusing on the particular solution to $\zeta$ which will be responsible for the peak leading to PBH production, $\zeta$ goes like the  square of the vector field which is Gaussian. Using $g$ for this Gaussian field,

\begin{align}
\zeta = g^2 - \langle g^2 \rangle \nonumber\\
\langle \zeta^2 \rangle = 2 \langle g^2 \rangle^2  \nonumber\\
\sigma^2 = \langle g^2 \rangle \nonumber\\
\sqrt{\frac{1}{2} \langle \zeta^2 \rangle} =  \sigma^2 
\end{align}

\noindent Plugging this in:

\begin{align}
\beta = erfc \left( \sqrt{\frac{1}{2} \left(\frac{\sqrt{2} \, \delta_c}{\sqrt{P_{\zeta}}   }  + 1 \right) }   \right)
\end{align}

\section{$\phi F \tilde{F}$ Model}

First we'll consider the contribution to the power spectrum from production of the vector field. This will be done for an arbitrary background inflaton potential $V(\phi)$. In the next section we can consider results from specific $V(\phi)$ models.  The Lagrangian is:

\begin{align}\label{eqr}
\mathcal{L} = - \sqrt{-g} \left( \frac{1}{2} g^{\mu \nu} (\partial_{\mu} \phi) (\partial_{\nu} \phi) + V(\phi) + \frac{1}{4} F_{\mu \nu} F^{\mu \nu} + \frac{\alpha}{4 } \phi F_{\mu \nu} \tilde{F}^{\mu \nu} \right)
\end{align}

\noindent We'll need the 0th order equations for $\phi$ and $A$ and the equation of motion of  $\delta \phi$ to find the power spectrum. Using conformal time with prime denoting $\frac{d}{d \tau}$, the 0th order equation of motion for $\phi$ is:

\begin{align}
\phi''+ 2 \frac{a'}{a} \phi' - \nabla^2 \phi + a^2 V_{\phi} = - a^2 \frac{\alpha}{4 } F_{\mu \nu} \tilde{F}^{\mu \nu}
\end{align}

\noindent The last term can be expanded using that $F_{\mu \nu} = \partial _{\mu} A_{\nu} - \partial_{\nu} A_{\mu}$ and $\tilde{F}^{\mu \nu} = \frac{1}{2} \frac{1}{\sqrt{-g}} \epsilon^{\mu \nu \alpha \beta} (\partial_{\alpha} A_{\beta} - \partial_{\beta} A_{\alpha})$, where $\epsilon$ is the standard Levi-Civita tensor. $\sqrt{-g} = a^4$ in conformal time.  We can choose to work in Coulomb gauge where $A^0 = 0$.

\begin{align}
\phi''+ 2 \frac{a'}{a} \phi' - \nabla^2 \phi + a^2 V_{\phi} = -  \frac{\alpha}{ a^2 }  \epsilon^{ i j k}  A_{i}^{'} \,  \partial_{j} A_{k} 
\end{align}

\noindent This can be rewritten in terms of the electric and magnetic fields $E$ and $B$ using:

\begin{align}
&E_i = - \frac{1}{a^2} A^{' }_i \\
&B_k = \frac{1}{a^2} (\nabla \times A)_k = \frac{1}{a^2} \epsilon_{ijk} \partial_i A_j
\end{align}

\begin{align}
\phi'' - \frac{2}{\tau}\phi' + a^2 V_{\phi} = a^2 \alpha  \langle E \cdot B \rangle
\end{align}

\noindent As is convention for this model, we define $\xi = \frac{\alpha \dot{\phi}}{2 H}$. The above can then be rewritten using $\langle E \cdot B \rangle = - \alpha (2 \times 10^{-4})  \frac{H^4}{\xi^4} e^{2 \pi \xi}$, eq. (\ref{eqeb}.) 

\begin{align}\label{eqt}
\phi'' - \frac{2}{\tau}\phi' + a^2 V_{\phi} =- \alpha (2 \times 10^{-4})  \frac{H^4}{\xi^4} e^{2 \pi \xi}
\end{align}

\noindent I'll also write this 0th order equation in efolding time as this is how it will get used in the numerical simulation. 

\begin{align}\label{equ}
\phi_{NN} + ( \epsilon - 3) \phi_N + \frac{1}{H^2} V_{\phi} = - \alpha (2 \times 10^{-4}) \frac{H^2}{\xi^4} e^{2 \pi \xi}
\end{align}

The 0th order equation of motion of the vector field $A$ from the Lagrangian, eq. (\ref{eqr}), is

\begin{align}
0 = - \partial_{\alpha} \partial^{\beta} A^{\alpha} + \partial_{\alpha} \partial^{\alpha} A^{\beta} - \partial_{\alpha} \left( \alpha \phi \epsilon^{\mu \nu \beta \alpha} \partial_{\mu} A_{\nu} \right)
\end{align}

\noindent Using Coulomb gauge where $A^0 = 0$, this reduces to the equations

\begin{align}\label{eqAfield}
0 = \left( \frac{\partial }{\partial \tau^2} - \nabla^2 - \alpha \phi' \nabla \times \right) A \nonumber\\
{\bf \nabla \cdot A} = 0
\end{align}

\noindent where $\tau$ is conformal time. We move A into momentum space, promote to an operator, and decompose into creation and annihilation operators:

\begin{align}\label{eqy}
\hat{A}(\tau, {\bf x}) = \int \frac{d^3 {\bf k}}{(2 \pi)^{3/2}} e^{i {\bf k \cdot x}} \hat{A}(\tau, {\bf k})
\end{align}

\begin{align}\label{eqz}
A^i (\tau, {\bf k})= \sum_{\lambda = \pm} \left( \epsilon^i_{\lambda}({\bf k}) \, u_{\lambda}(\tau, {\bf k}) \, \hat{a}_{\lambda}({\bf k}) + \epsilon^{* i}_{\lambda}(- {\bf k}) \,u^*_{\lambda}(\tau, - {\bf k}) \, \hat{a}^{\dagger}_{\lambda}(- {\bf k}) \right)
\end{align}

\noindent The equations of motion for the mode functions which solve eq. (\ref{eqAfield}) are:

\begin{align}\label{eqs}
0 = u^{''}_{\pm} + \left(k^2 \pm \frac{2 k \xi}{\tau} \right) u_{\pm}
\end{align}

Depending on the arbitrary sign choice of $\dot{\phi}$, one of the two mode functions will be exponentially enhanced while the other stays essentially in vacuum. I'll choose $ \phi$ to roll towards larger values making $\dot{\phi}$ and  therefore $\xi$ positive. The result is $u_+$ gets exponentially enhanced while $u_-$ does not. $u_-$ can then be ignored. This sign choice also determines the sign of $\langle E \cdot B \rangle$ appearing in eq. (\ref{eqt}).

Instead of writing the full solution of eq. (\ref{eqs}), a convenient approximation can be made. We're interested in $u_+$ when it's large and near its peak. When $\frac{1}{8 \xi} \ll |k \tau| \ll 2 \xi$, $u_+$ is well approximated by:

\begin{align}\label{eqx}
u_+ (\tau, k) \approx \frac{1}{\sqrt{2k} } \left( \frac{k}{2 \xi a H}  \right)^{\frac{1}{4}} e^{\pi \xi - 2 \sqrt{\frac{2 \xi k}{a H}}}
\end{align}

\noindent where $a$ is the scale factor.

\subsection{Finding the Perturbed Equation and Power Spectrum}

To find the power spectrum, we need the equation of motion for $\delta \phi$. Expanding the action gives: 

\begin{align}
S =& - \int d^4 x\, \sqrt{-g} \, \Big(\frac{1}{2} \partial ( \phi + \delta \phi)^2 + \frac{1}{4} (F + \delta F) (F + \delta F) + \frac{\alpha}{4} (\phi + \delta \phi) (F \tilde{F} + \delta (F \tilde{F}) ) + V(\phi) \nonumber\\
& + \delta \phi \cdot \frac{\partial V}{\partial \phi} + \frac{1}{2} \delta \phi^2 \cdot \frac{\partial^2 V}{\partial \phi^2} \Big)
\end{align}

\noindent Expanding the equations of motion to include all terms 2nd order in both $\phi$ and $A$ to account for the backreaction of $\phi$ on $A$ as well as the backreaction of $A$ on $\phi$ leads to equations difficult to solve. Instead many authors (\cite{Anber:2009ua, Barnaby:2011qe,            Linde:2012bt, Garcia-Bellido:2016dkw}) include only the most significant contribution to the backreaction on $A$ which comes from the dependence of $A$ on $\xi$ and therefore $\dot{\phi}$. This can be expressed through the partial derivative third term in the following expansion, using $- \frac{1}{4} F \tilde{F} = E \cdot B$:

\begin{align}
\frac{1}{4} \delta(F \tilde{F}) &= \frac{1}{4} \left( F \tilde{F} - \langle F \tilde{F} \rangle + \frac{\partial F \tilde{F}}{\partial \dot{\phi}}  \cdot \delta \dot{\phi} \right) \nonumber\\
&= - E \cdot B + \langle E \cdot B \rangle - \frac{\partial (E \cdot B)}{\partial \dot{\phi}} \delta \dot{\phi}
\end{align}

\noindent This produces the equation of motion of $\delta \phi$:

\begin{align}
 \delta \ddot{\phi} + \left(3  H - \alpha \frac{\partial (E \cdot B)}{\partial \dot{\phi}} \right) \delta \dot{\phi}  - \frac{1}{a^2} \nabla^2  \delta \phi +  V_{\phi \phi} \cdot \delta \phi   = \alpha \left(  E \cdot B - \langle E \cdot B \rangle   \right) 
\end{align}

Then we can use $\langle E \cdot B \rangle = - 2 \times 10^{-4} \frac{H^4 e^{2 \pi \xi}}{\xi^4}$ and $\xi = \frac{\alpha \dot{\phi}}{2 H}$ to find

\begin{align}
\frac{\partial \langle E \cdot B \rangle}{\partial \dot{\phi}_0} = \frac{\alpha \pi}{ H} \langle E \cdot B \rangle \left(1 - \frac{2}{\pi \xi} \right)
\end{align}

\noindent Plugging this into $\delta \phi$ equation gives:

\begin{align}
 \delta \ddot{\phi} + \left(3  H -  \frac{\alpha^2 \pi}{H} \langle E \cdot B \rangle \left(1 - \frac{2}{\pi \xi} \right) \right) \delta \dot{\phi}  - \frac{1}{a^2} \nabla^2  \delta \phi +  V_{\phi \phi} \cdot \delta \phi   = \alpha \left(  E \cdot B - \langle E \cdot B \rangle   \right) 
\end{align}

\noindent This can be rewritten by pulling out the $3H$ and replacing one of the $\alpha$'s using:  $\xi = \frac{\alpha \dot{\phi}}{2 H}$. Let

\begin{align}
\beta =& 1 - \frac{2 \pi \alpha \xi}{3 H \dot{\phi}} \langle E \cdot B \rangle \left(1 - \frac{2}{ \pi \xi} \right) \nonumber\\
 =& 1 + \frac{2 \pi}{3} (2 \times 10^{-4}) \frac{\alpha H^3}{\xi^3 \dot{\phi}} e^{2 \pi \xi}  \left(1 - \frac{2}{ \pi \xi} \right)
\end{align}

\noindent giving

\begin{align}
  \delta \ddot{\phi} + 3 H\, \beta \, \delta \dot{\phi} - \frac{1}{a^2} \nabla^2 \delta \phi +  \delta \phi V_{\phi \phi}  =  \alpha (E \cdot B - \langle E \cdot B \rangle    )
\end{align}

To estimate of the power spectrum in the region where backreaction is large where we're interested in $\delta \phi$ near horizon crossing, I will follow the simple approximation of \cite{Barnaby:2011qe, Linde:2012bt} in saying $\partial_0 \approx H$. This causes the 1st and 3rd terms to cancel. The equation of motion reduces to:

\begin{align}
3 H^2\, \beta \, \delta \phi +  \delta \phi V_{\phi \phi}  =  \alpha (E \cdot B - \langle E \cdot B \rangle    )
\end{align}

It can also be shown that the 2nd term is also much smaller than the first term. Comparing $3 H^2 \beta \gg V_{\phi \phi} $ is equivalent to comparing $\beta \gg \eta$ after using the Friedmann equation $H^2 = \frac{V}{3 M_P^2}$ and the slow roll parameter $\eta = \frac{V_{\phi \phi}}{V} M_P^2$.  $\beta$ starts at 1 in the non-backreaction regime and grows to $\gg 1$, so this is true.

\begin{align}
 3 H^2\, \beta \, \delta \phi \approx  \alpha (E \cdot B - \langle E \cdot B \rangle    )
\end{align}

We can then write the power spectrum in the backreaction regime using $\zeta = - \frac{H}{\dot{\phi}} \, \delta \phi$ and $P_{\zeta} = \frac{k^3}{2 \pi^2} |\zeta|^2$.  Let $\sigma_{EB}$ be the variance in $E \cdot B$, $\sigma^2 = \langle  (E \cdot B)^2 \rangle   - 2  \langle E \cdot B \rangle^2 + \langle E \cdot B \rangle^2$.  $\sigma_{EB} = 2 \times 10^{-4} \frac{H^4}{\xi^4} e^{2 \pi \xi}$, eq. (\ref{eqsigma}.) This gives the particular solution valid at late times when backreaction is significant:

\begin{align}\label{eqaa}
P_{\zeta\, p} = \frac{\alpha^2 \sigma_{EB}^2}{9 \beta^2 \dot{\phi}^2 H^2} 
\end{align}

\noindent  To this should be added the homogeneous solution, which dominates at early times. 


\begin{align}
 P_{\zeta} = P_{\zeta \, standard} +  P_{\zeta\, p}
\end{align}

\noindent where $P_{\zeta \, standard} $ is the homogeneous solution, the standard, single-field power spectrum, $P_{\zeta \, standard} = \frac{H^4}{4 \pi^2 \dot{\phi}^2}$. One could do a more precise calculation of $P_{\zeta}$ in the middle regime. \cite{Ballardini:2019rqh} provides a careful regularization scheme to handle the backreaction exactly. However, for large $\xi$, the approximations above work well, and it is the large $\xi$ regime that leads to significant PBH production.


\section{Inflaton Models  }

\subsection{Natural Inflation }

In the original natural inflation model, the inflaton is a pseudo Nambu-Goldstone boson in which a slightly broken shift symmetry is used to motivate the flatness of the inflaton's potential such that it satisfies the slow roll conditions. The generated pNGB potential is \cite{PhysRevLett.65.3233}

\begin{align}
V = \Lambda^4 \left(1 + \cos \left(\frac{\phi}{f} \right) \right)
\end{align}

It's common to define $\chi = \frac{\phi}{f}$. Inflation starts with $\chi$ between 0 and $\pi$, and $V$ decreases as $\chi$ rolls towards  larger positive values. Inflation must end with $\chi < \pi$ so inflation ends with $V$ positive. These considerations will determine what range of parameters are allowed. 

It looks like there are two free parameters $\Lambda$ and $f$. However, if you restrict yourself to using Planck's values for $n_s$ and $A_s$, this ends up reducing to one free parameter. There are further restrictions on the remaining free parameter such that there is a solution consistent with reasonable assumptions regarding reheating. 

I'll consider $f$ to be my one free parameter. First I'll give bounds on $f$ in the single field inflation model as a point of comparison. Below, I'll include backreaction from a $\phi F \tilde{F}$ interaction. As $f$ gets larger, the energy scale determined by $\Lambda$ gets larger and $\chi_k$, the value of $\chi$ at Planck's pivot, increases. This means the pivot location is further from the top of the hill. Also the length of inflation from Planck's pivot till inflation ends, what I'll call $N_k$, decreases. For extremely large $f$ in the hundreds of $M_P$, inflation still takes about 60 efolds, but occurs for a tiny range of $\chi$ values just less than $\pi$.

There is a lower bound on $f$. As $f$ decreases, $\chi_k$ and $\Lambda$ decrease such that the energy scale of inflation is smaller, and the inflaton starts closer to the top of the hill where the potential is flatter. One may suppose there is a minimum $f$  since $\chi$ can only start so close to the top of the hill and still give the correct $n_s$. This minimum $f$ is around $5.47 M_P$. However, even before that limit is reached, the requirement that the model satisfy reasonable assumptions about reheating is passed.  As $f$ decreases, the length of inflation increases. The longer inflation goes on, a larger range of wavelength and frequency scales are able to freeze out of the horizon  as measured from the Planck's pivot scale to the end of inflation.\footnote{Technically many more scales might have frozen out before the pivot scale since we don't know how long inflation lasted. However,  modes which froze out before CMB scales will be unobservable.} This translates into a wider range of scales which then have to reenter the horizon after inflation. This generally forces the average equation of state during reheating to be larger. Most would agree the average equation of state during reheating, $w_{re}$, is almost certainly less than 1 (corresponding to a kinetic energy dominated field) and is likely less than or equal to 1/3 (corresponding to radiation.)  Applying such assumptions gives a minimum value for $f$. For Planck's preferred $n_s$ and $A_s$ and for single field inflation with the above potential, the minimum value of $f$ consistent with $w_{re} \leq 1$ is about $6.5 M_P$ with a corresponding $\Lambda = .0052 M_P$. Raising $f$ lowers the minimum $w_{re}$ value which will give a solution. However, even for $f = 100 M_P$, the lowest $w_{re}$ which gives a solution (with logical positive values for the length of reheating, $N_{re}$, and the length of the radiation epoch, $N_{RD}$) is $w_{re} = 0.40$. However, if one supposes $n_s$ isn't exactly at its preferred value but is a little smaller, this allows for smaller values of $f$ without violating reasonable assumptions about reheating. I'll explore the effect of lowering $n_s$ below.

So far these bounds have assumed single field inflation, and including backreaction from a $\phi F \tilde{F} $ interaction will have a small effect on shifting these bounds. However, I wanted to start with the single field bounds as an initial benchmark comparison point. This gives roughly the range of $f$ and $\Lambda$ values one might be interested in.

\subsubsection{Calculating the Bounds on $f$}

To find these bounds, I'm going to use slow roll approximations. First I'll write an expression relating Planck's $n_s$ value and an assumed $f$ value to $N_k$. Starting from $N = \int H \, dt$, this can be written:

\begin{align}
N_k = \frac{1}{M_P^2} \int_{\phi_{end}}^{\phi_k} d \phi \, \frac{V}{V_{\phi}}
\end{align}

\noindent One can simplify to:

\begin{align}
N_k = \frac{f^2}{M_P^2} \ln \left(\frac{1- \cos \chi_{end}}{1 - \cos \chi_k} \right)
\end{align}

\noindent where end stands for the end of inflation. We can replace the $\chi_{end}$  using that when inflation ends, $\epsilon = 1$. This reduces to:

\begin{align}
\cos \chi_{end} = \frac{- 1 + \frac{M_P^2}{2 f^2}}{1 + \frac{M_P^2}{2 f^2}}
\end{align}

\begin{align}\label{eqc}
N_k = \frac{f^2}{M_P^2} \ln(\frac{2}{(1+\frac{M_P^2}{2 f^2})(1 - \cos \chi_k)})
\end{align}

\noindent Next I'll take the slow roll parameters $\epsilon = \frac{M_P^2}{2} \left(\frac{V_{\phi}}{V} \right)^2$ and $\eta = M_P^2 \frac{V_{\phi \phi}}{V}$  and plug these into the standard slow roll expression for $n_s$, $n_s = 1 - 6 \epsilon - 2 \eta$. Using the natural inflation potential, this gives:

\begin{align}\label{eqb}
n_s = 1 + \frac{M_P^2}{f^2} \frac{(-3 + \cos \chi_k)}{1 + \cos \chi_k}
\end{align}

I combined the $N_k(\chi_k)$ and $n_s(\chi_k)$ equations (equations \ref{eqb}, \ref{eqc}) to get $N_k(n_s)$:

\begin{align}\label{eqd}
N_k = \frac{f^2}{M_P^2} \ln \left(\frac{1- n_s + \frac{M_P^2}{f^2}}{ \left(1+ \frac{M_P^2}{2f^2} \right) \left(1 - n_s - \frac{M_P^2}{f^2} \right) } \right)
\end{align}

\noindent The $n_s( \chi_k)$ equation can be inverted to give $\chi_k (n_s)$:

\begin{align}\label{eqg}
\cos \chi_k = 1 - 2 \frac{\left(1-n_s - \frac{M_P^2}{f^2} \right)}{1 - n_s + \frac{M_P^2}{f^2}}
\end{align}

Then to get $\Lambda( f, n_s)$, I used a combination of the standard  inflation formulas: $r = 16 \epsilon$, $r = \frac{P_h}{A_s}$, and $P_h = \frac{2 H^2}{\pi^2 M_P^2}$ to get

\begin{align}\label{eqf}
\Lambda = \left( 6 \pi^2 M_P^4 A_s \frac{ \left(1 - n_s - \frac{M_P^2}{f^2} \right)}{1 + \cos \chi_k}  \right)^{\frac{1}{4}}
\end{align}

\noindent Thus once an $f$ value is chosen to test, equation \ref{eqd} gives $N_k$, the length of inflation, and equations \ref{eqg} and \ref{eqf} give $\Lambda$.  Then I numerically solved the inflaton equation of motion in efolding time:

\begin{align}\label{eqi}
\phi_{NN} + \frac{H_N}{H} \phi_N - 3 \phi_N + \frac{V_{\phi}}{H^2} = 0 
\end{align}

\noindent from the pivot scale $N_k$, eq \ref{eqd}, till the end of inflation, $N = 0$. As this is a 2nd order differential equation, this requires two initial conditions. I used an initial value for $\phi$ from eq \ref{eqg} and an initial value for $\phi_N$ from assuming slow roll held initially. The standard slow roll equation of motion in real time $3 H \dot{\phi} \approx - V_{\phi}$ becomes 

\begin{align}\label{eqj}
\phi_N \approx \frac{V_{\phi} M_P^2}{V}
\end{align}

\noindent in efolding time using $\dot{\phi} = - H \phi_N$.

The starting assumptions I'm using assume the inflaton starts on a hill at a small, positive value of $\chi$ and slowly rolls to large $\chi$ values. To find the minimum value of $f$ regardless of reheating considerations,  one finds if $f$ is too small, the above equation for $\chi_k$ (equation \ref{eqg}) predicts an imaginary value for $\chi_k$. 

To find the minimum value of $f$ using reheating considerations, I numerically calculated $H(N)$ and $V(N)$. I plugged these into reheating estimates for $N_{re}$ and $N_{ND}$. The following equations \ref{eqk} and \ref{eql} hold for $w_{re} \neq 1/3$. For a derivation of the following formulas see for example \cite{Dai:2014jja}.

\begin{align}\label{eqk}
N_{re} = \frac{4}{(1- 3 \, w_{re})} \left(61.55 - \ln \left(\frac{V^{\frac{1}{4}}(N=0)}{H_k} \right) - N_k \right)
\end{align}

\begin{align}\label{eql}
N_{RD} = \ln \left(\frac{H_k}{k_{pivot} } \right) - N_k - \ln \left( \frac{a_0}{a_{eq}} \right) - N_{re}
\end{align}

\noindent $k_{pivot}$ is Planck's pivot value $= 0.05 \frac{1}{\mbox{Mpc}} $.  If $w_{re} =1/3$, $N_{re}$ and $N_{RD}$ effectively blend together, but you can solve for their sum:

\begin{align}
N_{re} + N_{RD} = \ln \left(\frac{a_{eq}}{a_0} \right) - N_k  - \ln \left(\frac{k_{pivot}}{a_0 H_k} \right)
\end{align}

\noindent When $f$ is too small, the only values of $w_{re}$ which provide a solution (with $N_{re}$ and $N_{RD}$ both positive) require $w_{re} > 1$.

\subsubsection{Including the Vector Coupling}

Now suppose the background $\phi$ potential is the same, but let's  include the coupling between $\phi$ and the vector field.  Instead of considering bounds on the homogeneous equation of motion of the inflaton, I'll look at solutions including the coupling with the vectors which backreact on the inflaton, eq \ref{equ} and \ref{eqv}: 

\begin{align}\label{eqm}
\phi_{NN} + \left(\frac{H_N}{H} - 3 \right) \phi_N + \frac{V_{\phi}}{H^2} = - 2 \times 10^{-4} \alpha \frac{H^2}{\xi^4} e^{2 \pi \xi}
\end{align}

and 

\begin{align}\label{eqn}
H = \sqrt{ \frac{1}{2 g} \left( M_P^2 - \frac{1}{6} \phi_N^2 -  \sqrt{ \left( M_P^2 - \frac{1}{6} \phi_N^2 \right)^2 - 8 \frac{g}{6 V}} \right) }
\end{align}

with

\begin{align}\label{eqo}
g = \frac{1}{12 \sqrt{2} \pi^2 } e^{2 \pi \xi} \left(\frac{.005438}{\xi^3} + \frac{.01136}{2 \xi^5} \right)
\end{align}

\noindent I solved these equations numerically with the same boundary conditions used in the single field case. I started integrating from the pivot scale with the initial value of $\phi$ chosen to give the correct $n_s$ value at the pivot, equation \ref{eqg}.  For the starting value of $\phi_N$, I used the standard slow roll assumption  since vector production was minimal and slow roll assumptions were valid at the pivot scale, equation \ref{eqj}.  I also plugged in a  value of $\Lambda$ chosen to give the correct $A_s$,  eq. \ref{eqf}.

\begin{figure} 
\centering
    \includegraphics[width=14cm]{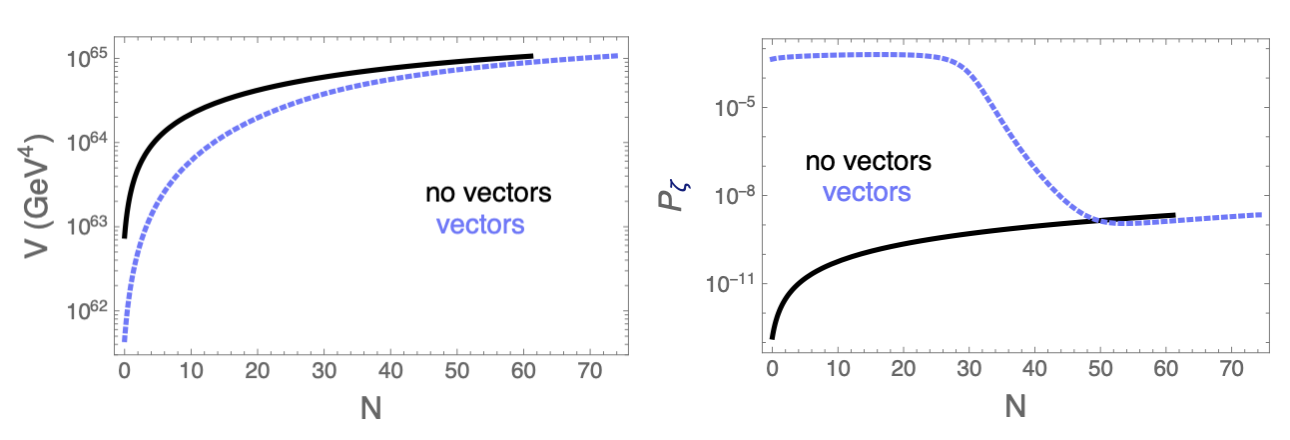}
\caption{The figures show the effect of adding the vector coupling. Both scenarios are plotted from Planck's pivot scale until inflation ends. Both use the same inflaton model with $f = 10\, M_P$ and $\Lambda = 1.83 \times 10^{16}$ GeV. $\xi_k =2.2$ or equivalently $\alpha = 1.67 \times 10^{-17}\, \mbox{GeV}^{-1}$  is used in the vector coupling case.}
\label{fig:natural_vector_no_vector}
\end{figure}

Figure \ref{fig:natural_vector_no_vector} examines the effect of including the $\phi F \tilde{F}$ coupling. Both scenarios are plotted starting at Planck's pivot scale and extending until inflation ends at $N = 0$. Both use the same inflaton model with $f = 10\, M_P$ and $\Lambda = 1.83 \times 10^{16}$ GeV. $\xi_k =2.2$ or equivalently $\alpha = 1.67 \times 10^{-17}\, \mbox{GeV}^{-1}$  is used in the vector coupling case. The $V$ shown is just the inflaton's potential energy. Energy from the inflaton sources production of the vectors, which are generated at an increasing rate during inflation. This slows the roll of the inflaton field. This is most pronounced at the end of inflation, where $V$ drops more rapidly in the no vector coupling scenario.  This also causes $N_k$, the length of inflation, to be longer with the vector coupling in place. In the example shown, $N_k$ increases from 61.1 to 74.2 when the vector coupling is added. 

Both power spectra start at Planck's observed value at the pivot. In the single field case, the power spectrum gradually decreases as $V$ decreases and the rolling of the inflaton increases as is standard. In the scenario with the vectors, the vectors are initially irrelevant, and this same pattern occurs. Then the vector production becomes increasingly significant causing $P_{\zeta}$ to rise.  Eventually backreraction of the vectors on $\phi$ becomes significant, slowing further increases in production of the vectors and $P_{\zeta}$ flattens out.  In the example shown, $P_{\zeta}$ peaks at $6.3 \times 10^{-4}$.  The purple dashed line including the vector coupling is the same purple dashed line which is the middle scenario in the following plots which example the effect of altering the key model parameters.

\begin{figure} 
\centering
    \includegraphics[width=14cm]{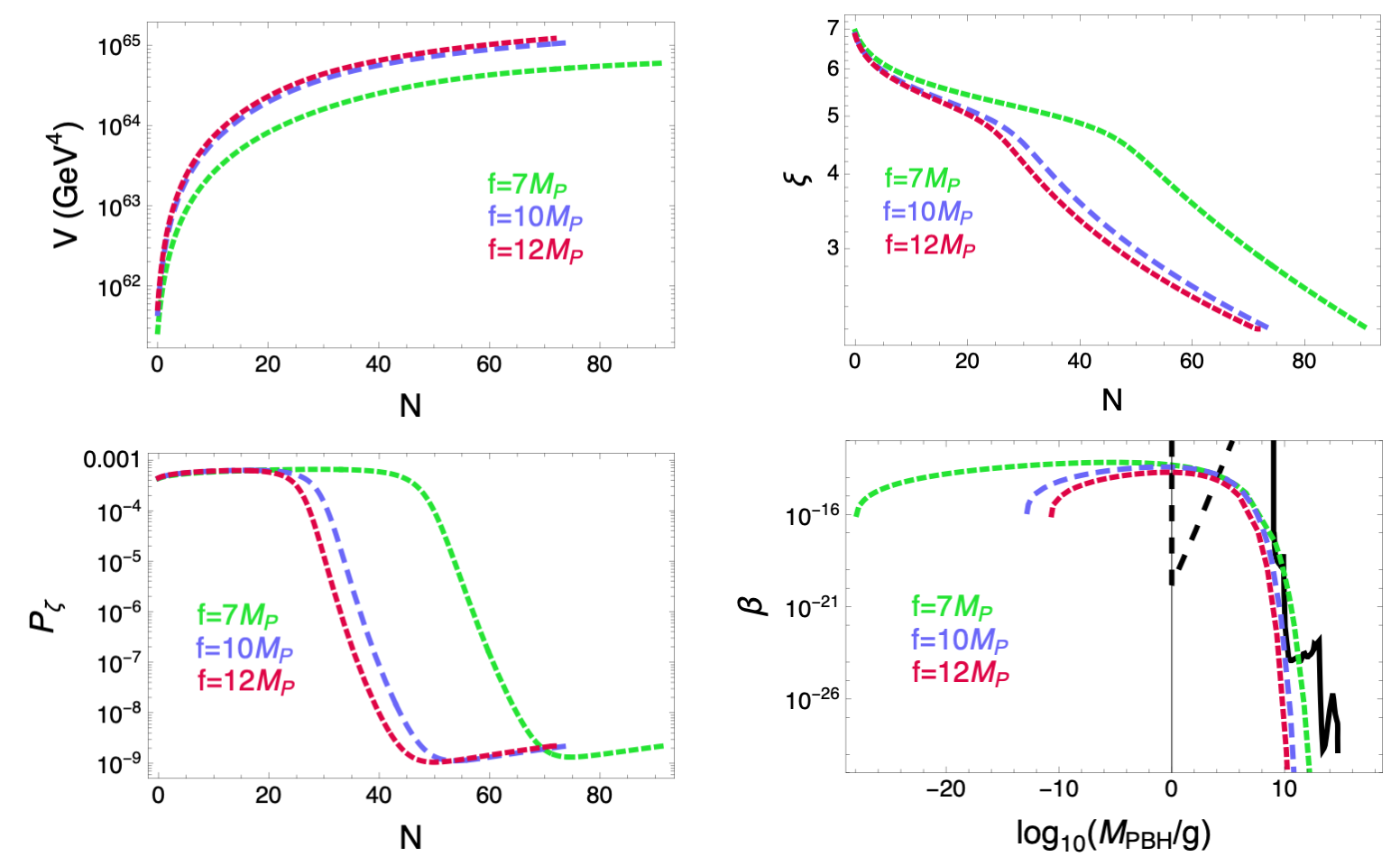}
\caption{The effect of changing $f$ while keeping all other parameters constant is examined. All three scenarios depicted use $\xi_k = 2.2$ and Planck's preferred values of $n_s$ and $A_s$. All have $\Lambda$ chosen to fit $A_s$ although this means $\Lambda$ is different for each scenario. The first three plots are shown as a function of efolding time $N$. They all start at Planck's pivot scale and count down to 0 when inflation ends.  The first plot shows $V$, the inflaton's potential energy. The second plot shows $\xi$ which is correlated to the rate at which vectors are produced. There are kinks in the $\xi$ figure from when backreaction of the produced vectors on the $\phi$ field becomes significant. $P_{\zeta}$ shows the total scalar power spectrum. It flattens in the backreaction regime. The final plot shows $\beta$, the fraction of the total energy density which collapsed into PBHs of the mass indicated.  The mass scales depicted here would all have evaporated before today. The three colored, dashed lines are predictions from the model.  The solid and dashed black lines shows constraints from \cite{Carr:2020gox}. The dashed bound on the left is only a possible bound from formation of stable Planck mass relics. At the relevant scales, the constraints on the right come from PBHs evaporating and affecting BBN. }
\label{fig:af}
\end{figure}

That's the overview, but more specifically we can look at the effect of changing specific model parameters. Below you'll see the effect of changing $f$, the coupling strength between $\phi$ and the vector field (displayed as the starting value of $\xi$), and $n_s$.  First let's examine the effect of changing $f$, the same parameter considered above which changes the background inflaton potential. The results are summarized in Figure \ref{fig:af}. Note in all three scenarios presented, I use Planck's preferred $A_s$ and $n_s$ values and the same starting value of $\xi = 2.2$. Note the results presented for different values of $f$ will shift if a different starting value of $\xi$ is chosen. In other words, a value of $f$ which doesn't work well won't necessarily never work, but might work if coupled with a different $\xi$. This effect of changing $\xi$ will be separately examined below. The starting value of $\xi$ is related to the coupling strength between the inflaton and the vectors through $\xi = \frac{\alpha \dot{\phi}}{2  H}$. However, note that $\dot{\phi}$ (and to a lesser extent $H$) is different in the three scenarios and so requiring they use the same starting value of $\xi$ means they all use different values of the coupling constant $\alpha$. The three values of the coupling constant $\alpha$ used were $\alpha = 1.57 \times 10^{-17}$, $1.67 \times 10^{-17}$, and $2.23 \times 10^{-17}$ GeV$^{-1}$  for $f = 7 M_P$, $f = 10 M_P$, and $f = 12 M_P$ respectively. In a similar way, all three scenarios have $\Lambda$ chosen to fit $A_s$ although this means $\Lambda$ is different for each scenario. $\Lambda = 1.41 \times 10^{16}$, $1.85 \times 10^{16}$, and $2.05 \times 10^{16} $ GeV for the $f = 7 M_P$, $10 M_P$, and $12 M_P$ scenarios.

The three lines for each $f$ start at Planck's pivot scale ($k = 0.05 \frac{1}{Mpc}$) and end when the slow roll parameter $\epsilon = 1$ (or equivalently when the equation of state reaches $1/3$) and inflation ends. Decreasing $f$ increases $N_k$, the length of inflation. Notice in the first three plots with $N$ on the $x$ axis, the $f = 7 M_P$ lines start at the largest $N$, meaning more efolds occur between the pivot scale and the end of inflation, $N_k= 91.6$. In contrast, $N_k= 74.2$ and 71.7 efolds in the $f = 10 M_P$ and $f = 12 M_P$ scenarios. 

Once $f$ is chosen, there is only one spot on the potential, one $\phi$ value, that will generate the correct $n_s$. This determines the location of the pivot. For smaller values of $f$, this occurs closer to the top of the hill of the potential where the inflaton rolls more slowly. This lengthens the number of efolds before inflation ends. See for example equation \ref{eqg}. 

The first plot of $V$, the inflaton's potential, shows the energy scale increasing as $f$ increases. Even though $\chi$ starts larger (further from the hilltop) in this case, $\Lambda$ is larger, leading to a larger $V$. See for example equation \ref{eqf}. For fixed $A_s$ and $n_s$, as $f$ gets larger, $\Lambda$ gets larger.

The second figure shows $\xi$. There is a kink where backreaction becomes significant which has the effect of slowing the growth of $\xi$. The lines for the three values of $f$ all start with $\xi = 2.2$ at the pivot scale. Since the pivot scale happens earlier, at larger $N$, in the $f = 7 M_P$ case, $\xi$ reaches larger values sooner at larger $N$, and the backreaction region is reached sooner. The third plot shows $P_{\zeta}$, the scalar power spectrum. All three lines start at the pivot scale at Planck's observed value. Initially $P_{\zeta}$ decreases like it normally does in single field inflation. At these early times, the inflaton is rolling down its potential and the energy density of vectors is still small. Then there is a marked increase in $P_{\zeta}$ from production of the vector field. Then backreaction kicks in, and $P_{\zeta}$ flattens out. 

The last figure shows $\beta$, indicating the mass scales at which significant numbers of PBHs are able to form for the three values of $f$. The $f = 10 M_P$ case was chosen specifically because it just hugs the BBN bound line (for $\xi_k = 2.2$) between what's ruled out and what might be observable in the future. The sharp edge to the $\beta$ lines on the left is a real feature. This left edge corresponds to PBHs formed from perturbations which froze out just as $N=0 $ at the end of inflation. There wouldn't be any perturbations further to the left.  

\begin{figure} 
\centering
    \includegraphics[width=6cm]{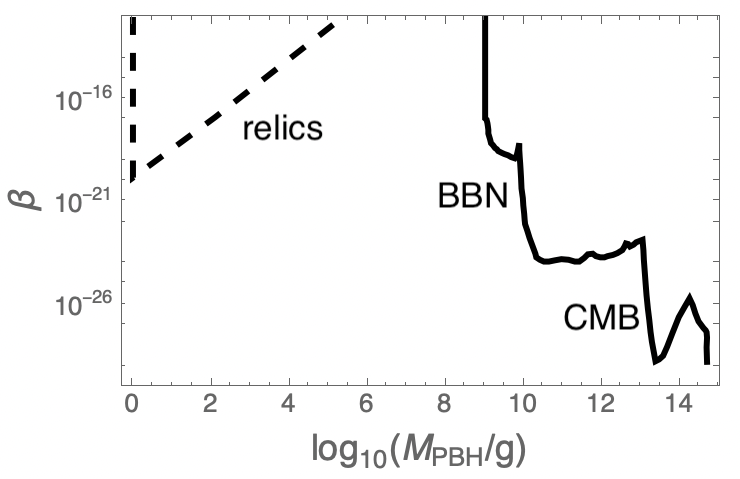}
\caption{The figure shows constraints on $\beta$ taken from  \cite{Carr:2020gox} labeled.  }
\label{fig:constraints}
\end{figure}

The black lines in this $\beta$ plot show bounds from \cite{Carr:2020gox}. These bounds are highlighted in Figure \ref{fig:constraints} with labels. At the mass scales considered, the PBHs are small enough that they should have evaporated. Evaporated particles off the PBHs will change the predictions for BBN and the CMB. The constraint labeled relics supposes these smallest PBHs didn't evaporate but formed stable Planck mass relics. I'm including results which violate the relics bound since it's not know if these relics would form. This is only a possible bound. This relics bound is shown as a dotted black line in Figure \ref{fig:af}.

The most significant effect of changing $f$ comes from the fact that the power spectrum stays large in the $f = 7 M_P$ case for a much longer time, and this translates into a much wider range of mass scales over which significant numbers of PBHs form. The maximum mass scale at which potentially observable PBHs form is roughly the same in the three cases. The main difference is how far production extends to even smaller mass scales. 

\begin{figure} 
\centering
    \includegraphics[width=6cm]{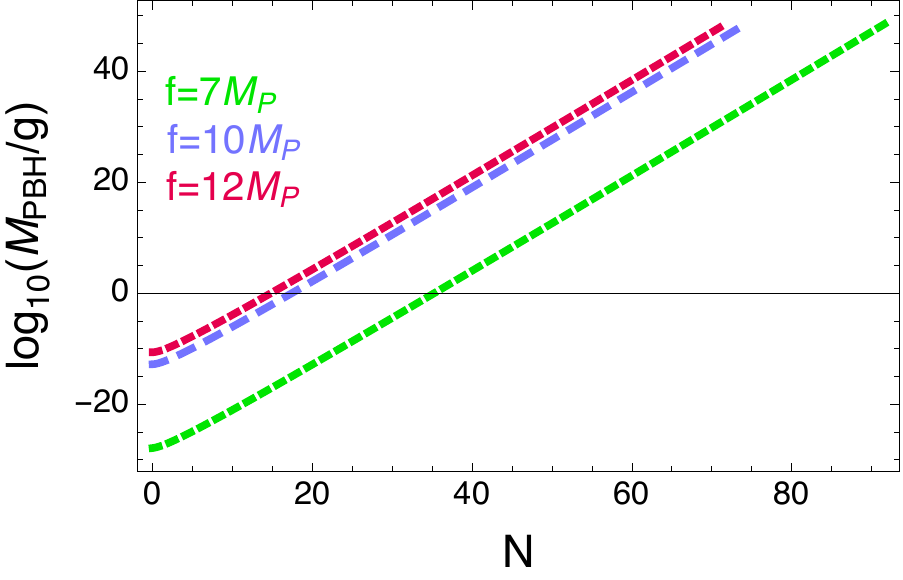}
\caption{The same three scenarios from the previous graph are shown. The relationship between the mass of the generated PBHs and the time during inflation when the corresponding perturbation exited the horizon is shown. }
\label{fig:nm}
\end{figure}

It's interesting that the maximum PBH masses are so similar when the corresponding $N$ values are so different.  Perturbations which form earlier correspond to larger PBHs, so one might expect the $f = 7 M_P$ scenario to produce larger PBHs than the others. However, there is a kind of competing effect, see Figure \ref{fig:nm}. For fixed $N$, the smaller $f$ scenario produces smaller mass PBHs. Therefore although the smaller $f$ scenario creates large perturbations at earlier times (larger $N$), those perturbations generate smaller mass PBHs than if the larger $f$ scenarios produced PBHs at those same $N$ values. The result is that the largest mass PBHs produced in the three scenarios are of similar size. To explain why, note equation \ref{eq3} states $M_{PBH}(N) \propto (\frac{a_0}{a_{end}})^2$ which is $\propto e^{2(N_{re} + N_{RD})}$. The most important term in $N_{re} + N_{RD}$ which varies between scenarios is the $- N_k$, see eq. \ref{eqw}. The longer inflation in the $f= 7 M_P$ scenario leads to smaller $N_{re} +N_{RD}$ and correspondingly smaller  $M_{PBH}(N)$.

One might look at the plot of $f = 7 M_P$ and corresponding $N_k = 91.6$, be reminded that many inflation models predict $N_k$ around 60, and question if this is allowed. The longer inflation puts strain on reheating. Reheating will have to last longer, occur at a larger equation of state, or both to get all of those extra frequency/ wavelength modes which exited the horizon to reenter after inflation ends. The smallest $w_{re}$ which accommodates a solution (with $N_{re}$ and $N_{RD}$ positive) for the 3 cases is: 2.3, 0.87, and 0.78, for $f = 7 M_P$, $f = 10 M_P$, and $f = 12 M_P$ respectively.

\begin{figure} 
\centering
    \includegraphics[width=14cm]{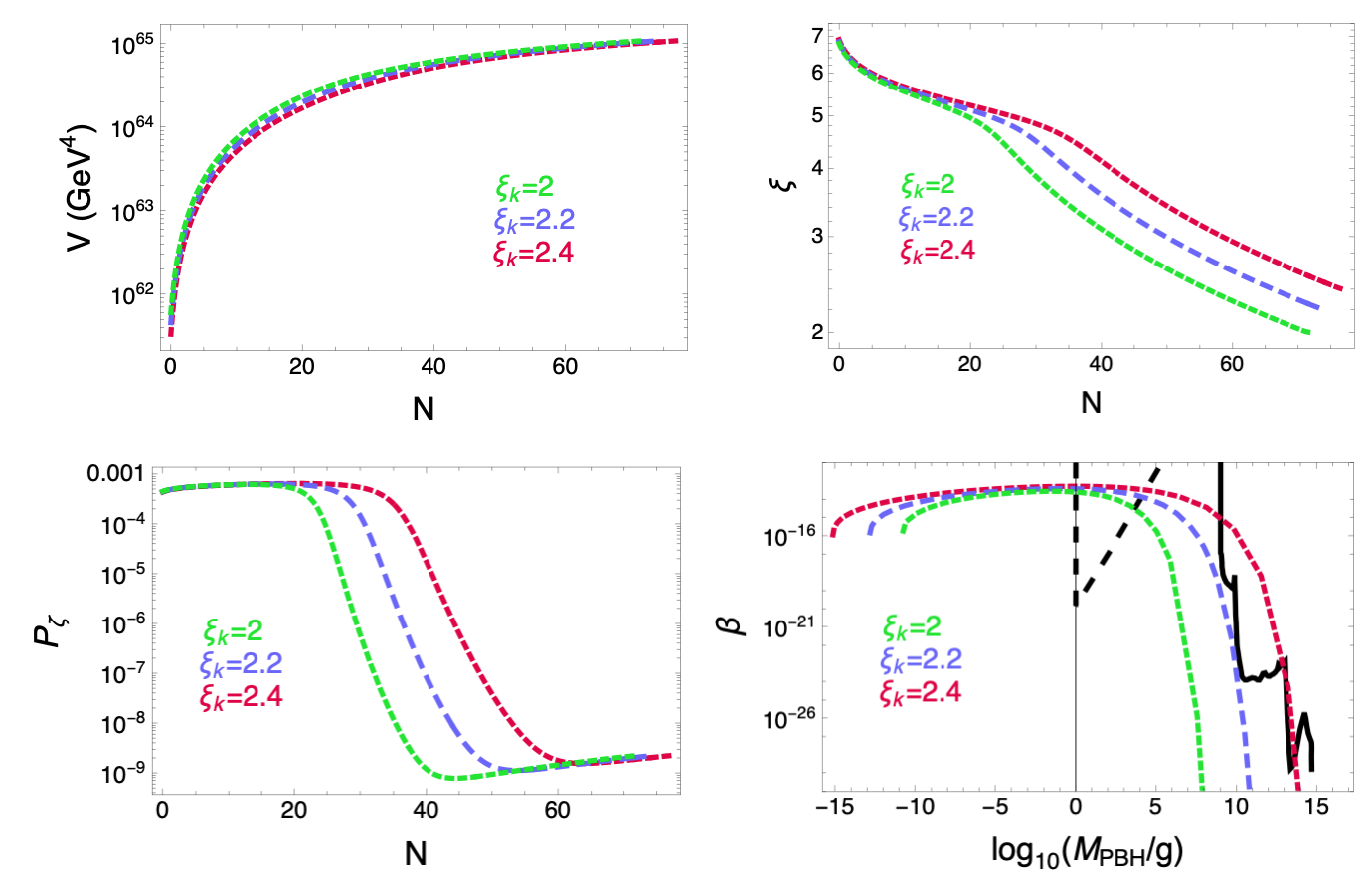}
\caption{The figures show the effect of changing $\xi_k$, the value of $\xi$ at the pivot. All three scenarios depicted use $f = 10 M_P$ and Planck's preferred values for $A_s$ and $n_s$. The first three plots are shown as a function of efolding time $N$. They all start at the pivot scale and count down to 0 when inflation ends. The first plot shows $V$, the inflaton's potential energy. The second plot shows $\xi$ which correlates to production of the vector field. $P_{\zeta}$ shows the total scalar power spectrum. There is a feature in each of the plots of $\xi$ and $P_{\zeta}$ when backreaction becomes significant. The final plot shows $\beta$, the fraction of the energy density which collapsed into PBHs of the mass indicated.  The three colored, dashed lines are predictions from the model. The solid and dashed black lines shows constraints from \cite{Carr:2020gox}. The dashed bound on the left is only a possible bound from formation of stable Planck mass relics.  }
\label{fig:natural_xi}
\end{figure}

The next parameter I examined varying is the value of $\xi$ at the pivot, $\xi_k$. Larger $\xi_k$ translates to a larger coupling between $\phi$ and the vector field. The results are shown in Figure \ref{fig:natural_xi}. The middle purple line is the same as in the previous plot, using $\xi_k = 2.2$. All three scenarios in this figure use $f = 10 M_P$ and Planck's observed values for $n_s$ and $A_s$. $\Lambda$ is chosen to fit $A_s$; $\Lambda = 1.85 \times 10^{16}$ GeV in each scenario. 

Notice there is little difference in the inflaton's potential $V$ for the three scenarios since I'm not changing the inflaton's potential directly but only its coupling to the vectors. Until backreaction becomes significant, the inflaton's equation of motion and therefore its potential is largely unchanged. One of the small differences is that the larger $\xi_k$ is, the larger the effect of backreaction slowing the roll of the inflaton's potential, and this lengths inflation somewhat. $N_k = 71.7$, 74.2 and 77.0 for $\xi_k = 2.0$, $\xi_k = 2.2$, and $\xi_k = 2.4$ respectively.  When $\xi_k$ starts larger, larger numbers of vectors are produced earlier, and the backreaction regime is reached sooner. One can see the kink in the $\xi$ and $P_{\zeta}$ plots occurring earlier for the larger $\xi_k$ scenario. As seen in the previous figure, a longer period with a larger $P_{\zeta}$ translates into a wider range of mass scales over which PBHs can form/ a wider range of mass scales where $\beta$ is large. There is increased PBH production for the large $\xi_k$ case both at the higher and lower ends of the mass range.  The smallest $w_{re}$ which accommodates a solution (with $N_{re}$ and $N_{RD}$ positive) for the three cases is: 0.76, 0.87, and 1.0, for $\xi_k = 2.0$, $\xi_k = 2.2$, and $\xi_k = 2.4$ respectively.

\begin{figure} 
\centering
    \includegraphics[width=14cm]{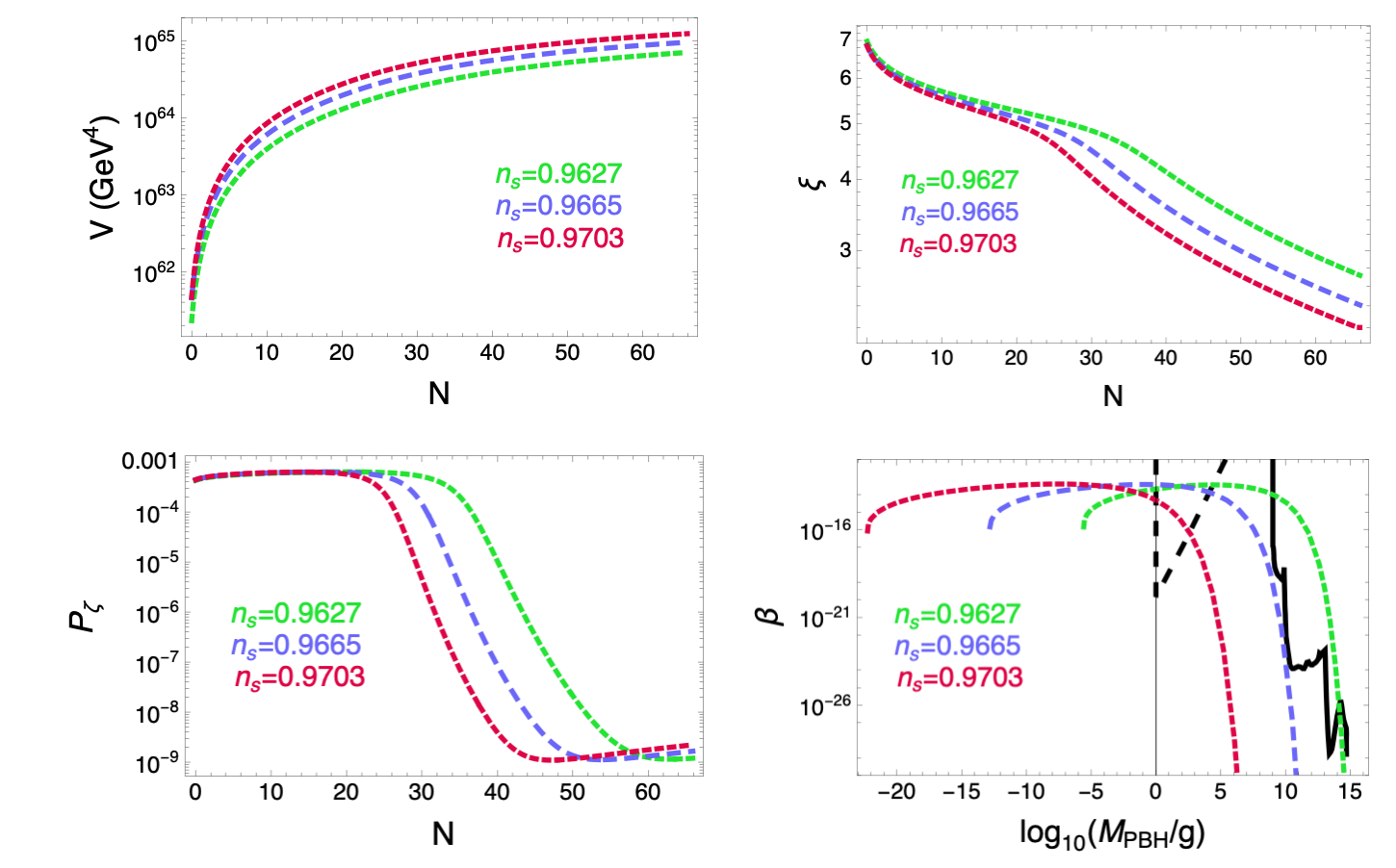}
\caption{As in the previous figures, the middle, purple line is unchanged. All three scenarios depicted use $f = 10 M_P$, $\xi_k = 2.2$ and Planck's preferred value for $A_s$. The middle, purple scenario uses Planck's preferred value for $n_s$ while the other two use the $1 \sigma$ bounds. Each plot starts at Planck's pivot scale and continues until inflation ends. The solid and dashed black lines in the forth plot shows constraints from \cite{Carr:2020gox}. }
\label{fig:natural_ns}
\end{figure}

I've been using Planck’s preferred value of $n_s$, but there are error bars, and that preferred value always shifts a little as more data is acquired. It's interesting to examine how the parameter space would change if the preferred $n_s$ value is modified a little up or down. See Figure \ref{fig:natural_ns}. The middle purple scenario is the same as in the previous plots. All three scenarios use Planck's preferred $A_s$, $f = 10 M_P$, and $\xi_k = 2.2$. The higher and lower $n_s$ values are Planck's $1 \sigma$ bounds. In each scenario $\Lambda$ is chosen to make $A_s$ fit Planck's preferred value, but this leads to a different $\Lambda$ in each scenario. $\Lambda = 1.96 \times 10^{16}$, $1.85 \times 10^{16}$, and $1.73 \times 10^{16}$ GeV for the $n_s = .9627$, .9665, and .9703 scenarios.

Changing $n_s$ changes where the pivot is. A larger $n_s$ corresponds to a spot higher up on the potential, closer to the top of the inflaton's hill with a more slowly rolling inflaton and smaller slow roll parameters. Because of this, inflation lasts longer in this scenario. $N_k = 85.3$ in the larger $n_s$ case compared to $N_k = 74.2$ and $N_k = 65.9$ in the preferred and smaller $n_s$ scenarios. The longer inflation with a flatter potential also corresponds to a smaller energy scale shown in the $V$ plot of the figure and from eq. \ref{eqf}.

Since larger $n_s$ means inflation takes longer, and I'm starting $\xi_k$ at the same value of 2.2 in each scenario, larger $n_s$ means $\xi$ reaches a larger value sooner. The backreaction region is reached sooner, and $P_{\zeta}$ stays large for a longer time. This translates into a wider peak in the $\beta$ plot. PBHs form over a wider range of mass scales. However, notice PBHs are able to form at larger masses in the smaller $n_s$ scenarios.  The large $N_k$ in the large $n_s$ scenario means $N_{re}+N_{RD}$ is much smaller (eq. \ref{eqw}) and this means $M_{PBH}(N)$ is also much smaller for a given $N$ (eq. \ref{eq3}). Even though $P_{\zeta}$ grows large sooner, the resultant PBHs are smaller.

Again longer inflation pushes up the average equation of state of reheating. The smallest $w_{re}$ which accommodates a solution for the three cases is: 0.545, 0.865, and 1.57, for $n_s = 0.9627$, $n_s = 0.9665$, and $n_s = 0.9703$ respectively.

In conclusion, including the vector coupling will slow the rolling of the inflaton and prolong the length of inflation. All the scenarios considered would violate the bound of production of stable Planck mass relics if those would actually form, but that's not known. Assuming that bound isn't real, the right parameters could produce PBHs which evaporate and leave a signature on BBN parameters. $f = 10 M_p$ and $\xi_k = 2.2$ would just hug the current limit. Or for example, raising $f$ while lowering $\xi_k$ a small amount could also produce PBHs which hit the same bound.

\subsection{Hilltop Inflation}

The hilltop potential was introduced in \cite{LINDE1982389} and has continued to be one of the models which can match CMB data well. I'll consider the quartic hilltop model:

 \begin{align}
V = M^4 \left(1- \left(\frac{\phi}{\mu} \right)^4 \right) 
\end{align}

As long as the exponent is positive, as the name implies, inflation starts on the top of a hill with convention usually dictating the inflaton slowly rolls towards larger positive values. It's common to define $\chi = \frac{\phi}{\mu}$ in which case $\chi$ starts close to 0 and increases but always stays less than 1 so that $V$ remains positive.

Similar to natural inflation, it looks like there are two free parameters, $\mu$ and $M$. However, if one restricts themselves to using Planck's observed $A_s$ and $n_s$, this reduces to one free parameter, which I'll take to be $\mu$. Increasing $\mu$ increases the energy scale of inflation and increases $\chi_k$, pushing the location of the pivot further down from the top of the hill. This allows inflation to take less efolds from the pivot till inflation ends, meaning smaller $N_k$.  There are restrictions on how small $\mu$ can be since smaller $\mu$ leads to longer inflation and $N_k$ can't be arbitrarily long. The restriction comes from the fact that the more efolds between the pivot and the end of inflation, the larger the equation of state during reheating needs to be to allow all those frequency modes which froze out of the horizon to reenter. In the single field, no vector coupling case, the smallest $\mu$ which allows for a solution with $w_{re} \leq 1$ is $ \mu = 9.1 M_P$. The corresponding $M$ is $M \geq 6.95 \times 10^{15}$ GeV.  The minimum $\mu$ which gives a solution for $ w_{re}$ between 0 and 1/3 is $19 M_P$, corresponding to $M \geq 1.24 \times 10^{16}$ GeV. Although the vector coupling will alter these restrictions, this gives benchmark values for comparison. As was true for natural inflation, lowering $n_s$ pushes the pivot further down the hill towards larger $\chi_k$ values, and this makes it possible to find solutions at smaller $\mu$ values.

\subsubsection{Calculating the Bounds on $\mu$}

The standard slow roll parameters $\epsilon= \frac{M_P^2}{2} \left(\frac{V_{\phi}}{V} \right)^2$ and $\eta = M_P^2 \frac{V_{\phi \phi}}{V}$ for the quartic hilltop potential give:

\begin{align}
\epsilon = \frac{8 M_P^2}{\mu^2} \left(\frac{\chi^3}{1 - \chi^4} \right)^2
\end{align}

and

\begin{align}
\eta = - \frac{12 M_P^2}{\mu^2} \frac{\chi^2}{(1 - \chi^4)}
\end{align}

These can be plugged into the standard slow roll equation for $n_s$: $n_s = 1 - 6 \epsilon + 2 \eta $. Using Planck's observed $n_s$, this gives an equation fixing $\chi_k$, $\chi$ at the pivot, for a chosen $\mu$. For example, in the scenarios explored below I'll focus on $\mu = 30 M_P$, which then correlates to $\chi_k =  0.723$.

Next fitting to Planck's $A_s$ sets $M$. We can use the standard formula for the tensor spectrum, $P_h=\frac{2 H^2}{\pi^2 M_P^2 }$, and the tensor to scalar ratio $r$ to write: $r = \frac{2 H^2}{\pi^2 M_P^2 A_s}$. Using the standard slow roll expression $r = 16 \epsilon$ and the Friedmann equation, $H^2 = \frac{V}{3 M_P^2}$, gives:

\begin{align}
M^2 = \frac{8 \sqrt{3} \pi M_P^3 \sqrt{A_s}}{\mu} \frac{\chi_k^3}{(1- \chi_k^4)^{3/2}}
\end{align}

\noindent Once $\mu$ is chosen and we've found the corresponding $\chi_k$, this gives the corresponding $M$. For example, $\mu = 30 M_P$ gives $M = 1.15496 \times 10^{16}$ GeV. To find solutions in the single field case, I numerically solved the inflaton's equation of motion, equation, \ref{eqi}, using the hilltop potential. I integrated from the pivot scale to the end of inflation. The initial conditions at the pivot are the $\phi$ value which gives the correct $n_s$ and a $\phi_N$ which assumes slow roll at the pivot scale using equation \ref{eqj}. I used equations \ref{eqk} and \ref{eql} to make predictions for the length of reheating.

\subsubsection{Including the Vector Coupling}

\begin{figure}
\centering
    \includegraphics[width=14cm]{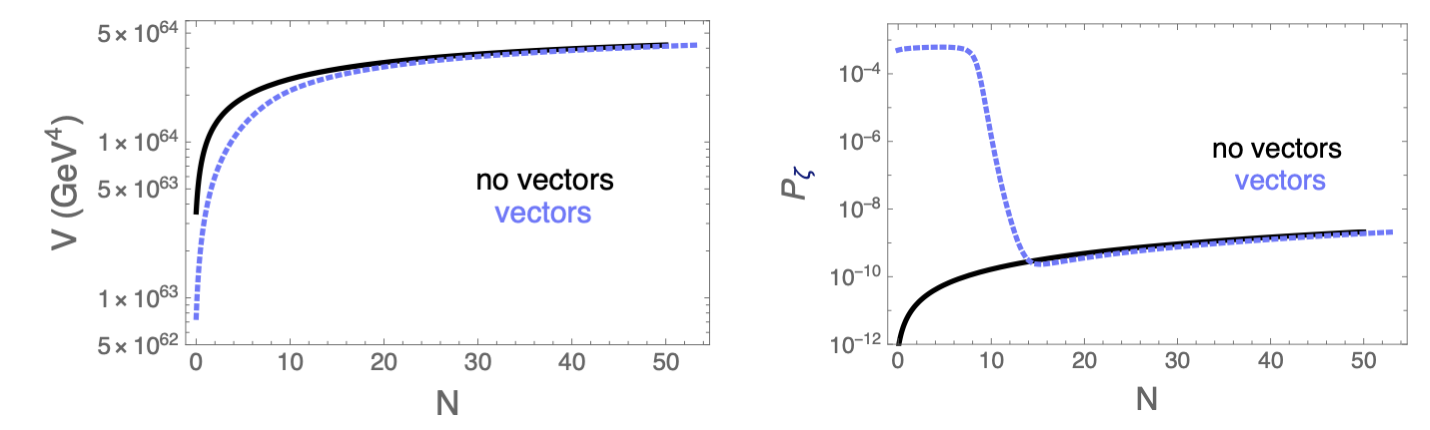}
\caption{The figure examines the effect of including the vector coupling while using the same background inflaton model parameters for both scenarios, $\mu = 30 M_P$ and $M = 1.55 \times 10^{16}$ GeV. Both scenarios are depicted starting at Planck's pivot and continuing until inflation ends at $N=0$. In the vector coupling scenario, $\xi_k = 1.16$  which corresponds to $\alpha = 1.38 \times 10^{-17}$ GeV$^{-1}$. }
\label{fig:hillvecnovec}
\end{figure}

Figure \ref{fig:hillvecnovec} examines the effect of including the vector coupling. The lines for both scenarios start at Planck's pivot scale and extend until inflation ends. The background inflation model is the same in both cases with $\mu = 30 M_P$. To agree with Planck's $n_s$, this fixes $\chi_k = 0.723$. To get Planck's $A_s$, this fixes $M = 1.55 \times 10^{16}$ GeV. I fixed the vector coupling $\alpha = 1.38 \times 10^{-17}$ GeV$^{-1}$ in the vector case to generate $\xi_k = 1.16$. This will be the central example shown in purple in the subsequent plots examining the effect of varying the model parameters.  

One of the effects of adding the vectors is some energy is passed from the rolling of the inflaton into producing the vector field. Inflation is prolonged from 50 efolds to 53.6 efolds. The power spectrum decreases as normal in both cases initially. Then the power spectrum shoots up due to production of the vectors when they're included, and then growth slows again when backreaction becomes significant. $P_{\zeta}$ peaks at $6.17 \times 10^{-4}$ in the example shown. 

To find the numerical solution with the vectors included, I  solved the inhomogeneous equation of motion for the infalton in efolding time, equations \ref{eqm}, \ref{eqn} and \ref{eqo}. The differential equation is integrated from the pivot scale until inflation ends with initial conditions set at the pivot scale. The initial conditions are the same for the scenarios with and without the vector coupling as backreaction on the inflaton is irrelevant initially.

\begin{figure} 
\centering
    \includegraphics[width=14cm]{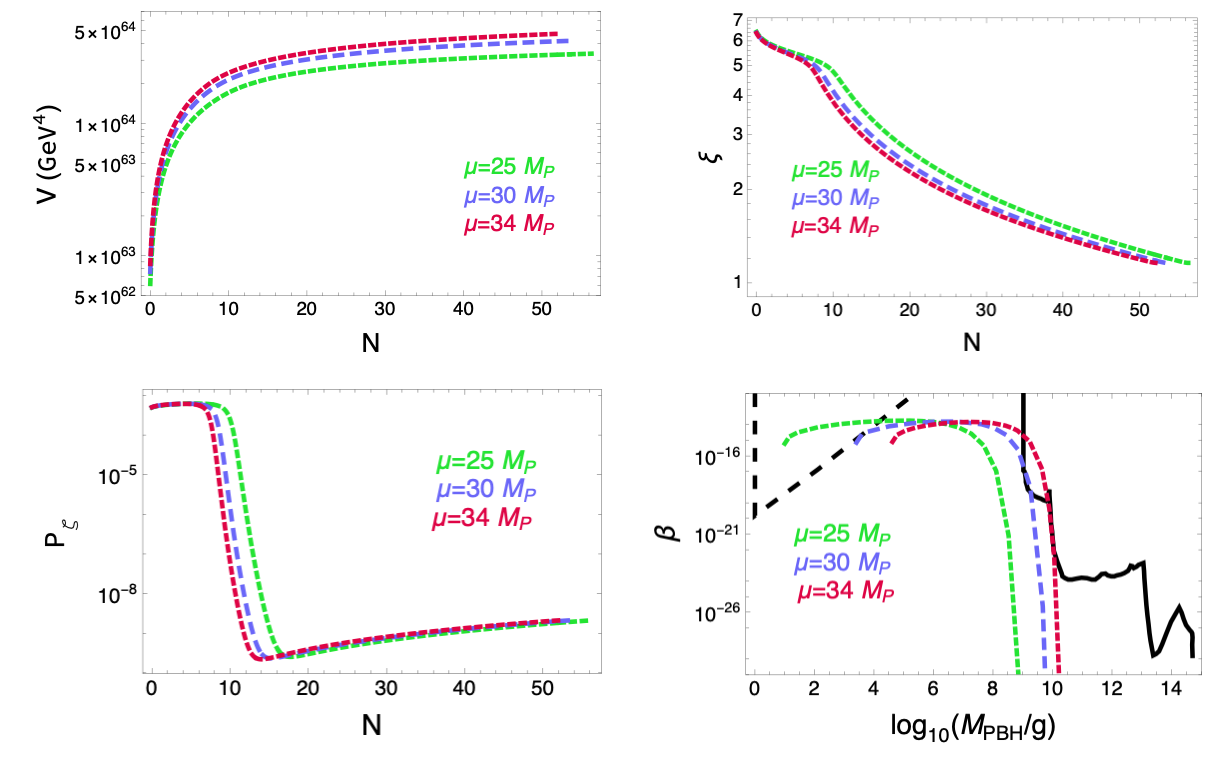}
\caption{The effect of changing $\mu$ while keeping all other parameters constant is examined. All three scenarios depicted use $\xi_k = 1.16$ and Planck's preferred values of $n_s$ and $A_s$.   The first three plots are shown as a function of efolding time $N$. They start at Planck's pivot  and count down to 0 when inflation ends.  The first plot shows $V$, the inflaton's potential energy. The second plot shows $\xi$ which is correlated to the rate at which vectors are produced. There are features in the $\xi$ figure from when backreaction of the produced vectors on $\phi$ becomes significant. $P_{\zeta}$ shows the total scalar power spectrum. It flattens in the backreaction regime. The final plot shows $\beta$, the fraction of the total energy density which collapsed into PBHs of the mass indicated. The mass scales depicted here would all have evaporated before today. The three colored, dashed lines are predictions for the model.  The black lines shows constraints from \cite{Carr:2020gox}. The dashed, black bound on the left is only a possible bound from formation of stable Planck mass relics. At the relevant scales the constraints on the right come from PBHs evaporating and affecting BBN.}
\label{fig:hillmu}
\end{figure}

Below I'll consider the effect of changing three of the key model parameters $\mu$, $\xi_k$, and $n_s$. The effect of altering $\mu$ is examined in Figure \ref{fig:hillmu}.  In each scenario presented, I use a starting $\xi_k = 1.16$ and Planck's preferred $n_s$ and $A_s$. $\xi$ relates to the coupling constant $\alpha$ through $\xi = \frac{\alpha \dot{\phi}}{2  H}$. However, as $\dot{\phi}$ and $H$ are different in the three scenarios, $\alpha$ must likewise be different in order to generate the same $\xi_k$. $\alpha = 1.54 \times 10^{-17}$ GeV$^{-1}$, $1.38 \times 10^{-17}$ GeV$^{-1}$, and $1.30 \times 10^{-17}$ GeV$^{-1}$ for the three scenarios $\mu =  25 M_P$, $30 M_P$, and $34 M_P$. In a similar way, $M$ is fit to Placnk's $A_s$ in each case, and this requires a different $M$ for each $\mu$. $M = 1.43 \times 10^{16}$ GeV, $1.55 \times 10^{16}$ GeV, $1.63 \times 10^{16}$ GeV for the three scenarios.

Changing $\mu$ changes the background inflaton model, and so you can see the changes in the inflaton's potential $V$. Increasing $\mu$ increases the energy scale and shortens the length of inflation. The three lengths of inflation depicted are 56.3, 53.6, and 52.2.  The smaller $\mu$ is, the smaller $\chi_k$ is, meaning the pivot scale is occurring closer to the top of the hill where the potential is flatter.

All the $\xi_k$ start at the same value of 1.16 at the pivot. However, since inflation lasts longer in the $\mu = 25 M_P$ case, $\xi_k$ is able to reach larger values sooner, further from the end of inflation. This likewise means that $P_{\zeta}$ is able to achieve larger values at larger values of $N$. This wider peak in $P_{\zeta}$ then translates into a wider $\beta$ peak, a wider range of mass values for which $\beta$ is large. As in the natural inflation plots, the solid and dashed black lines in the $\beta$ plot are bounds on PBH production from \cite{Carr:2020gox}. The dashed bound on the left is only a possible bound from formation of stable Planck mass relics. At the relevant scales, the constraints on the right come from PBHs evaporating and affecting BBN.  $\mu = 25 M_P$ crosses the bound on the left, however, this doesn't mean it's ruled out since it's not know if these Planck mass relics would form or not. For fixed $\xi_k$, increasing $\mu$ narrows the mass range over which PBHs form and shifts both the maximum and minimum mass over which they form towards larger mass values.  The sharp termination to the $\beta$ lines on the left is a real feature. This left edge corresponds to PBHs formed from perturbations which froze out just as $N=0 $ at the end of inflation. There wouldn't be any perturbations further to the left. All three scenarios presented have reheating solutions compatible with $w_{re}$ between 0 and 1/3.

It might seem counterintuitive that the larger $\mu$ scenario in which $P_{\zeta}$ only grows larger at the smallest $N$ still leads to the larger mass PBHs as seen in the $\beta$ figure.  If I instead plotted $\beta$ as a function of $N$, $\beta$ would grow larger earlier in inflation, at larger $N$ values in the smaller $\mu$ scenarios. The explanation is that,  $M_{PBH}(N)$ is much smaller at a given $N$ in the $\mu= 25 M_P$ case. The function $M_{PBH}(N)$ is exponentially sensitive to $N_{re}  + N_{RD}$, which is smaller for smaller $\mu$. 

\begin{figure}
\centering
    \includegraphics[width=14cm]{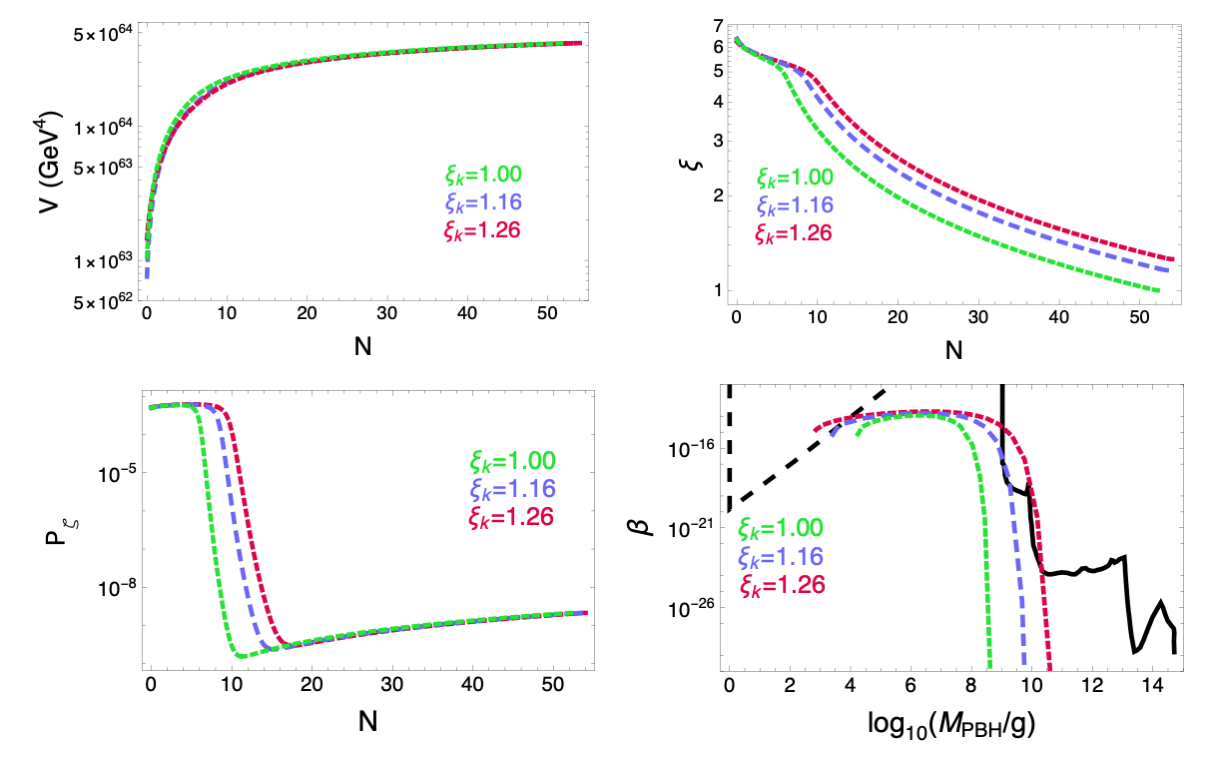}
\caption{The figures show the effect of changing $\xi_k$, the value of $\xi$ at the pivot. All three scenarios depicted use $\mu = 30 M_P$ and Planck's preferred values for $A_s$ and $n_s$. The first three plots are shown as a function of efolding time $N$. They all start at the pivot scale and count down to 0 when inflation ends. The first plot shows $V$, the inflaton's potential energy. The second plot shows $\xi$ which is correlated to the rate at which vectors are produced. $P_{\zeta}$ shows the total scalar power spectrum.  The final plot shows $\beta$, the fraction of the total energy density which collapsed into PBHs of the mass indicated.  The three colored, dashed lines are predictions for the model. The black lines shows constraints from \cite{Carr:2020gox}. }
\label{fig:hillxi}
\end{figure}

\begin{figure}
\centering
    \includegraphics[width=14cm]{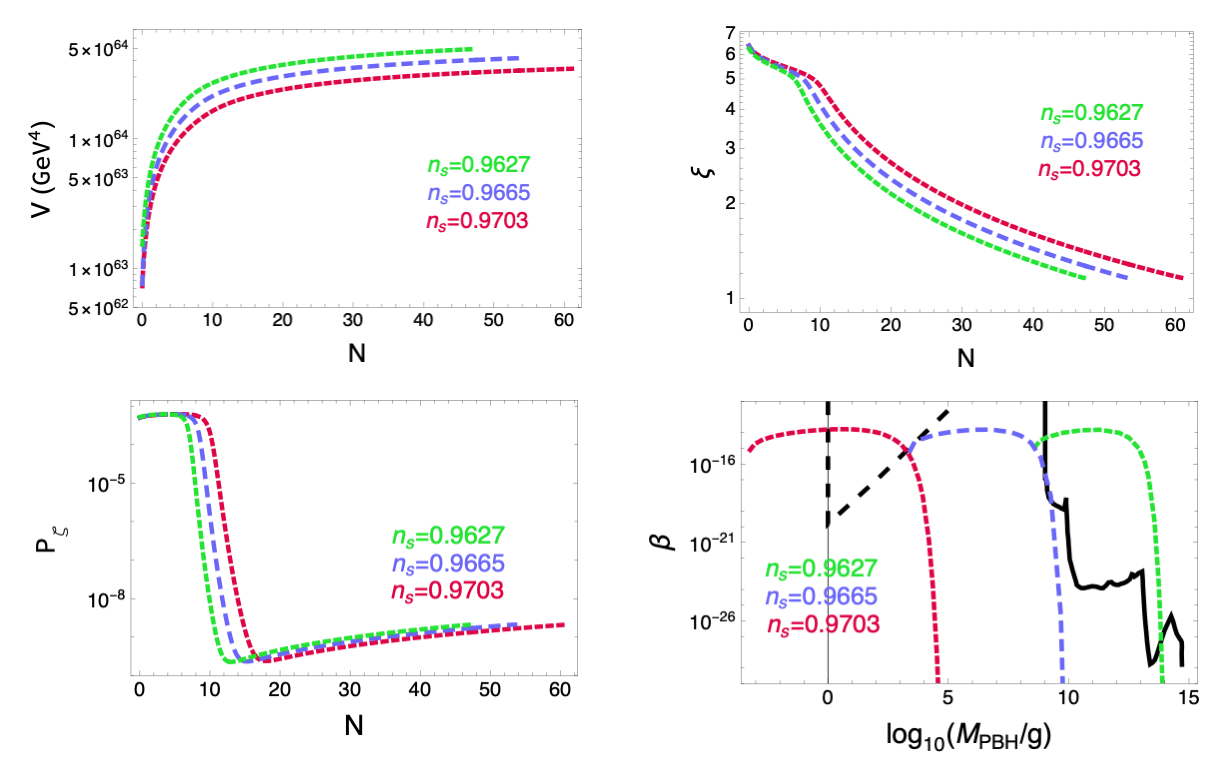}
\caption{The figures exaimine the effect of changing $n_s$. The middle, purple line is the same as in the previous two figures. All three scenarios use $\mu = 30 M_P$, $\xi_k = 1.16$, and $A_s$ at the preferred value. The three scenarios show Planck's central $n_s$ value and the $1 \sigma$ bounds. Each plot starts at Planck's pivot scale and continues until inflation ends. The three colored, dashed lines are predictions for the model. The black lines in the last figure shows constraints from \cite{Carr:2020gox}.}
\label{fig:hillns}
\end{figure}

As in the natural inflation plots, the central purple line is the same in Figures \ref{fig:hillmu}, \ref{fig:hillxi}, and \ref{fig:hillns}. It's chosen for the fact that its parameters just hug the bound set by BBN constraints. The second parameter I examined varying is the starting value of $\xi_k$ at the pivot scale. The effect is shown in Figure \ref{fig:hillxi}. The background inflaton model is the same in each scenario with $\mu = 30 M_P$. $M$ and the starting value of $\chi$ are fit to Planck's observed values of $n_s$ and $A_s$ such that $\chi_k = 0.723$ and $M = 1.55 \times 10^{16}$ GeV in each scenario. Since the background models are the same, the three $V$ plots are nearly identical. The only thing separating them is the vector production amount, and that only becomes significant late in inflation. Starting $\xi$ larger at earlier times means that $\xi$ has more time to grow exponentially. This leads to a larger backreaction effect of the vectors on $\phi$ and this slows the rolling of $\phi$, and inflation lasts longer. The length of inflation from the pivot scale till inflation ends, $N_k$, is 52.5, 53.6, and 54 for $\xi_k =1.00$, 1.15, and 1.26. Larger $\xi_k$ leads to $P_{\zeta}$ growing larger earlier and staying larger for a longer span of $N$ values. This leads to a wider peak in $\beta$, meaning PBHs form over a wider range of mass scales. Larger $\xi_k$ leads to PBH production at both larger and smaller mass scales. There are reheating solutions compatible with $w_{re}$ between 0 and 1/3 for all three scenarios.

The last parameter I examined varying is $n_s$. Although Planck has measured $n_s$, the measured value shifts a little as new data is collected, and it's interesting to see how small changes in $n_s$ will affect results. This is shown in Figure \ref{fig:hillns}. In each case $\mu = 30 M_P$, $\xi_k = 1.16$, and Planck's observed value of $A_s$ is used.  Larger $n_s$ is closer to the scale independent $n_s = 1$ case, so larger $n_s$ correlates to a pivot higher up on the potential where the potential is flatter. $\chi_k = 0.736$, 0.723, and 0.707 for the $n_s = 0.9627$, 0.9665, and 0.9703 scenarios. Larger $n_s$ also correlates to a smaller energy scale for inflation as seen in the $V$ plot. The flatter potential leads to a longer length of inflation. $N_k = 47.4$, 53.6, and 61.2 for the three scenarios. As inflation lasts longer in the larger $n_s$ scenario, and $\xi_k$ is starting at the same value, $\xi_k$ reaches larger values at larger $N$. This leads to $P_{\zeta}$ become large at larger $N$. This generates a wider peak in $\beta$, meaning PBH production over a wider range of mass scales. However, the most notable change in $\beta$ is the maximum and minimum mass scale at which PBHs form is substantially larger for smaller $n_s$ values. Even though $P_{\zeta}$ becomes larger at larger $N$ for larger $n_s$, the mass of PBHs formed at that scale is much smaller in the larger $n_s$ case. Long inflation leads to small $N_{re} +N_{RD}$, see eq. \ref{eqw}, which leads to small $M_{PBH}(N)$, see eq. \ref{eq3}.

Both the central and smaller $n_s$ values accommodate reheating solutions with $w_{re}$ between 0 and 1/3. However, the larger $n_s$ value is only compatible with $w_{re} \geq 0.46$.  In addition to inflation being longer in the larger $n_s$ case, reheating is shorter. $N_{re} + N_{RD}$ which is independent of $w_{re}$ is 67, 53.6, and 53.1 for the three $n_s$ values. The combination of longer inflation and shorter $N_{re} + N_{RD}$ means the equation of state during reheating has to be larger for all the extra frequency modes which froze out during inflation to have a chance to reenter the horizon.

\section{Conclusion}

Some of the black holes observed by LIGO have features unexpected from astrophysical black holes. Notably they've found binary partners with unexpectedly large masses and isotropically aligned spins.  This has increased interest in the possibility of PBHs.  Evidence from PBHs could give information about $P_{\zeta}$ at a drastically different frequency scale from the CMB. This could be extremely helpful in understanding inflation. 

Various groups have considered ways of generating a peaked power spectrum which becomes large enough to source PBHs but becomes small at late times in order to not violate the strong constraints from evaporated PBHs on BBN and the CMB. There has been less attention paid to the possibility of producing low mass PBHs from an increase in $P_{\zeta}$ at later times closer to the end of inflation which would evade these constraints from the other end.

Two background inflaton models, natural inflation and quartic hilltop inflation, were considered, paired with a $\phi F \tilde{F}$ vector coupling.  These background models were chosen because of how well they fit current Planck data. Vectors are produced in increasing numbers as inflation progresses, with production eventually slowing as backreaction on the inflaton becomes significant. The production of these vectors generates a peak in $P_{\zeta}$ which can later source PBHs during the subsequent radiation dominated era. 

In general, adding the vector coupling slows the rolling of the inflaton and increases the length of inflation. This lengthening can be quite substantial. However, the longer inflation goes on, the larger the equation of state of reheating will have to be in order that all the modes which freeze out during inflation are able to reenter the horizon after inflation ends. Applying restrictions like $w_{re} \leq 1$ or $\leq 1/3$ puts upper limts on the coupling strength and lower limts on the model parameters $f$ (natural inflation) and $\mu$ (hilltop inflation). 

There are three parameters considered which determine the amount of PBHs generated: the background inflation model parameters, the coupling strength to the vectors, and potentially $n_s$ if this gets modified by future CMB data. These parameters will determine the width and time at which a peak in $P_{\zeta}$ forms. The longer $P_{\zeta}$ stays large, the wider the range of mass scales over which PBHs are able to form. The closer the peak in $P_{\zeta}$ is to the end of inflation, the earlier after inflation those frequency modes will reenter the horizon and source PBHs, and the smaller those PBHs will be.  Small PBHs will evaporate and affect BBN and CMB predictions leading to strong constraints. However, if the peak in $P_{\zeta}$ occurs late enough, it's possible to hug current BBN bounds, allowing for the possibility of future detection without violating current limits.

Both background inflation models have two free parameters. However, both reduce to one free parameter if you require a fit to Planck's $n_s$ and $A_s$. Increasing $f$ (natural inflation) or $\mu$ (hilltop inflation) increases the energy scale and decreases the length of inflation. When inflation is shorter, $\xi$ and therefore $P_{\zeta}$ only become large close to the end of inflation. This means $P_{\zeta}$ is large for a shorter period of time and the resultant peak in $\beta$ is more narrow. PBHs are produced over a more narrow range of mass scales. This doesn't necessarily mean that the mass of the produced PBHs will be smaller. Higher energy shorter inflation correlates with a longer reheating and radiation dominated period and larger mass PBHs forming at a particular $N$ value during inflation. In the hilltop scenario, larger $\mu$ produced the largest mass PBHs, even though the perturbations which sourced them formed closer to the end of inflation. The larger the coupling between the vectors and the inflaton, tested through the starting value of $\xi$, the faster $P_{\zeta}$ grows large, and the wider the range of PBH masses sourced.

Which parameters works best is sensitive to $n_s$. The larger $n_s$ is, the flatter the potential at the pivot scale, and inflation will be at lower energy and last longer. For the same starting $\xi_k$ value, this leads to a wider peak in $P_{\zeta}$ with PBHs forming over a wider range of mass scales. However, the masses will be relatively low. 

\appendix

\section{Cutoff such that a PBH would have evaporated by today}


The evaporative lifetime of a black hole in years is given by \cite{Hawking:1974rv}:

\begin{align}
\tau_{evap} = \frac{G^2\, M^3}{\hbar\, c^4} \approx 10^{63}  \, \left(\frac{M_{PBH}}{M_{sun}} \right)^3
\end{align}

\noindent We want to ask the question, when would a PBH form such that it would just be evaporating today. This gives a minimum mass of black holes that would still be around. We'll set the evaporation time equal to the elapsed time between when the PBH formed and today. This  is given by integrating the first Friedmann equation:

\begin{align}
t = \frac{1}{H_0} \int_a^{a_0} da \frac{1}{a \sqrt{\Omega_r (\frac{a_0}{a})^4 + \Omega_m (\frac{a_0}{a})^3 + \Omega_{\Lambda}} }
\end{align}

\noindent However, we'll find that the time of formation is so early, this is nearly the age of the universe, $1.38 \times 10^{10}$ years.

We can use eq. \ref{eq8} for $M_{PBH}$ to write a function for the evaporation scale, $k_{evap}$, independent of inflation model. 

\begin{align}
\frac{k_{evap}}{a_0} = \left( \frac{ 10^{63} \, [yr]}{\tau_{evap}} \right)^{\frac{1}{6}} \, \left(\gamma \, \frac{1.26 \times 10^{-6} \, [GeV]}{ M_{sun} \, g^{1/6}} \right)^{\frac{1}{2}}
\end{align}

\noindent Assuming $ \gamma = 0.2$ and $g = 10.75$, as used above, and using $M_{sun} = 1.12 \times 10^{57}$ GeV, this gives a model independent estimate of $\frac{k_{evap}}{a_0}  = 8.0 \times 10^{-24}$ GeV. 

This can then be related to a time during inflation by finding the correspondence between when a PBH forms and the efolding $N$ when the perturbation responsible first froze out of the horizon. The frequency scales $k$ will be the same. When that mode reached horizon size, $k = a(N) \, H(N) $. Using conventions such that $N$ counts down to 0 at the end of inflation when $a = 1$, $a$ during inflation is given by $a = e^{-N}$, and $\frac{a_0}{a_{end}} = 3388 \, e^{N_{re}+N_{RD}} $. The length of reheating and the radiation epoch, $N_{re} + N_{RD}$ will be inflation model specific. 

\begin{align}
(8 \times 10^{-24}) \, 3388 \, e^{N_{re}+ N_{RD}}  = e^{-N} \, H(N)
\end{align}

\noindent This gives $N_{evap} = 12$ for hilltop inflation with $\mu = 30 M_P$ and $N_{evap} = 24$ for natural inflation with $f = 10 M_P$, both using Planck's values for $n_s$ and $A_s$. The big difference between models comes from the $e^{N_{re}} e^{N_{RD}}$ where even a small difference in $N_{re} + N_{RD}$ gets amplified.  One might then think this leads to a strong sensitivity to a reheating model, but it actually doesn't. Changing assumptions about reheating, changing an estimated $w_{re}$, will change the individual values of $N_{re}$ or $N_{RD}$ but won't change the sum $N_{re} + N_{RD}$.

\section{Finding $\sigma_{EB}$}

$\sigma_{E \cdot B}$ is required to evaluate the power spectrum. 

\begin{align}
\sigma^2 = \langle ( E \cdot B)^2 \rangle - \langle E \cdot B \rangle^2
\end{align}


\noindent Since the expressions are long, I'll separately calculate each term.

\noindent Using $E  = - \frac{A'}{a^2}$ and $B = \frac{1}{a^2} \nabla \times A$:

\begin{align}
\langle ( E \cdot B)^2 \rangle  = \frac{1}{a^8} \langle A^{'}_i \epsilon_{ilm} ( \nabla_l A_m) A^{'}_j \epsilon_{j ab} \nabla_a A_b \rangle
\end{align}

\noindent The $A$'s are moved into momentum space:

\begin{align}
\hat{A}^j({\bf x}) = \sum_{\lambda = \pm} \int \frac{d^3 {\bf k}}{(2 \pi)^{3/2}} e^{ i {\bf k \cdot x}}  \left( \epsilon^j_{\lambda}(\bf{k}) \, u_{\lambda}(\tau, \bf{k}) \, \hat{a}_{\lambda}(\bf{k}) +\epsilon^{* j}_{\lambda}(-\bf{k}) \, u^*_{\lambda}(\tau, -\bf{k})\, \hat{a}^{\dagger}_{\lambda}(-\bf{k})   \right)
\end{align}

\noindent  I'll only keep the $\lambda = +$ terms as they grow exponentially. 

\begin{align}
\langle ( E \cdot B)^2 \rangle  =& -\frac{1}{a^8 (2 \pi)^6} \epsilon_{ilm} \epsilon_{j ab} \int d^3 k \int d^3 k' \int d^3 p \int d^3 p'  e^{i k \cdot x} e^{ i k' \cdot x} e^{ i p \cdot x} e^{ i p' \cdot x}  \nonumber\\
 &\langle A^{'}_i(k)  k'_l A_m(k') A^{'}_j(p)  p^{'}_a A_b(p') \rangle
\end{align}

\noindent The non-zero terms will be of the form $\langle a a a^{\dagger} a^{\dagger} \rangle +\langle a a^{\dagger} a a^{\dagger} \rangle$.


\begin{align}
\langle ( E \cdot B)^2 \rangle =& - \frac{1}{a^8 (2 \pi)^6} \epsilon_{ilm} \epsilon_{j ab} \int d^3 k \int d^3 k'  \,\epsilon_{+ i} (k) \, u'  (k)  \Biggl(     k'_l \,   \, \epsilon_{+ m}(k') \, u(k')  \nonumber\\
 &  \left( \epsilon^*_{+ j}(k') \, u^{' *}(k') \epsilon^*_b(k) (- k_a) u^*(k) + \epsilon^*_{+ j}(k) u^{' *}(k) \epsilon^*_{+ b} (k') u^*(k') (- k^{'}_a) \right) \nonumber\\
& +  k_l \,  k^{'}_a \, \epsilon^*_{+j}(-k') u^{*'}(-k')  \epsilon_{+m}(-k) \, u(-k) \, \epsilon^*_{+b}(k') u^*(k') \Biggl)
\end{align}

Let the above $= T_1 + T_2 + T_3$ for the three terms above. As the expressions are long, we can solve for each term separately. I'll start with $T_1$ and $T_2$ together as they have the same angular integral. We'll also use the polarization vector identities: $(k \times \epsilon_{\pm}(k))_i = \mp i |k| \epsilon_{i \pm}(k)$ and  $(k \times \epsilon^*_{\pm}(k))_i = \pm i |k| \epsilon_{i \pm}^*(k)$. 



\begin{align}
T_1 + T_2 &=  \frac{1}{a^8 (2 \pi)^6}  \int d^3 k \int d^3 k' \, |k'| \Biggl( -  ( \epsilon_{+i}(k') \cdot  \epsilon_i(k) ) (\epsilon^*_j(k) \cdot  \epsilon^*_j(k')) \, u'  (k)   \, u(k')  \nonumber\\
 & \cdot (  |k| \, u^{' *}(k')\,  u^*(k) +  |k'|\, u^{' *}(k)\, u^*(k') ) + |k| \,  u'(k)  u^{*'}(-k') u(-k)  u^*(k') \Biggl)
\end{align}


\noindent Choosing $\hat{k}$ to be $\hat{z}$ and simplifying the polarization vectors gives: $( \epsilon_{+i}(k') \cdot  \epsilon_i(k) ) (\epsilon^*_j(k) \cdot  \epsilon^*_j(k')) = \frac{1}{4} (1 - \cos \theta')^2$. Then one can solve the angular integrals.


\begin{align}
T_1 + T_2 =& - \frac{2^4 \pi^2}{3 a^8 (2 \pi)^6}  \int d k \int d k'   \, u'  (k)   \, u(k')   (  |k|^3 |k'|^3 \, u^{' *}(k')\,  u^*(k) + |k|^2 |k'|^4 \, u^{' *}(k)\, u^*(k') ) 
\end{align}

Next I plug in for the mode functions, $u$ and $u'$. 

\begin{align}
T_1 + T_2 =& - \frac{e^{4 \pi \xi}}{3 \cdot 2^4 \pi^4 a^8} \Biggl( \left( \int d k   \, k^3 e^{-4 \sqrt{-2 \xi k \tau} }  \left(1 - \frac{1}{4 \sqrt{-2 \xi k \tau}} \right) \right)^2  \nonumber\\
&+  \int d k \int d k'  \, k^{5/2} \, k'^{7/2}    \left(1 - \frac{1}{4 \sqrt{-2 \xi k \tau}} \right) e^{-4 \sqrt{-2 \xi k \tau}}  e^{-4 \sqrt{-2 \xi k' \tau}}  \Biggl)
\end{align}

\noindent I'll integrate over the range over which the approximation for the mode functions is valid, $\frac{1}{8 \xi} < -k \tau < 2 \xi$.

\begin{align}
T_1 + T_2 =& - 3.03 \times 10^{-8} \, \frac{e^{4 \pi \xi} H^8}{\xi^8 }
\end{align}

 Next I'll simplify $T_3$.

\begin{align}
T_3 =& - \frac{1}{a^8 (2 \pi)^6} \epsilon_{ilm} \epsilon_{j ab} \int d^3 k \int d^3 k'  \,\epsilon_{+ i} (k) \, u'  (k) \,  k_l \,  k^{'}_a \, \epsilon^*_{+j}(-k') u^{*'}(-k')  \epsilon_{+m}(-k) \, u(-k) \, \epsilon^*_{+b}(k') u^*(k') 
\end{align}



\noindent The polarization vectors can be simplified noting $\epsilon_+(k) \cdot (k \times \epsilon_+(-k)) = - i |k|$ and $\epsilon_+^*(-k') \cdot (k' \times \epsilon_+^*(k')) = - i |k'|$.

\begin{align}
T_3 =\frac{1}{a^8 (2 \pi)^6} \int d^3 k \int d^3 k' \, |k| \, |k'|  u'(k)  u^{*'}(-k') u(-k)  u^*(k')
\end{align}

\noindent Note $u(k) = u(-k)$ and $u = u^*$, so the $k$ and $p$ integrals are the same. 

\begin{align}
T_3 =\frac{1}{a^8 (2 \pi)^6}  \left(  \int d^3 k\,  |k|   u'(k)  u(k)  \right)^2
\end{align}

\noindent We'll find this is the same as $\langle E \cdot B \rangle^2$ and so both terms will cancel in calculating $\sigma^2$. The angular integral just gives $4 \pi$. Then we can plug in for $u$ and $u'$.


\begin{align}
T_3 =  \frac{e^{4 \pi \xi}}{ 2^4 \pi^4 a^8 } \left(  \int d  k \, k^3   e^{- 4 \sqrt{-2 \xi k \tau}} \left(1 - \frac{1}{4 \sqrt{-2 \xi k \tau}} \right) \right)^2
\end{align}

\noindent Again we integrate over the region where our approximations are valid, $\frac{1}{8 \xi} < -k \tau < 2 \xi$.


\begin{align}
T_3 =  4.35 \times 10^{-8} \, \frac{H^8 \, e^{4 \pi \xi}}{\xi^8 }  
\end{align}

\subsection*{Calculation of $\langle E \cdot B \rangle$ }


\begin{align}
\langle E \cdot B \rangle = \left\langle - \frac{A^{'}_i}{a^2} \cdot \frac{1}{a^2} \epsilon_{ilm} \partial_l A_m \right\rangle
\end{align}


\noindent Again we expand out the operators, moving them into momentum space.



\begin{align}
\langle E \cdot B \rangle =  -\frac{i}{a^4 (2 \pi)^3} \epsilon_{i lm}   \int d^3 k \int d^3 k' e^{ i k \cdot x} e^{ i k' \cdot x} k^{'}_l \langle \epsilon_{+ i}(k) u'(k) \hat{a}(k) \epsilon^*_{+ m}(-k') u^*(-k') \hat{a}^{\dagger}(-k')   \rangle
\end{align}



\noindent The polarization vectors can be simplified using the identities: $(k \times \epsilon^*_{\pm}(k))_i = \pm i |k| \epsilon_{i \pm}^*(k)$ and $\epsilon_{i\pm}(k) \cdot \epsilon^*_{i \pm} (k) = 1$.



\begin{align}
\langle E \cdot B \rangle = - \frac{1}{a^4 (2 \pi)^3}   \int d^3 k \, |k|  u'(k)   u^*(k) 
\end{align}



\noindent Then we plug in for the mode functions $u$ and $u'$.


\begin{align}
\langle E \cdot B \rangle = - \frac{e^{2 \pi \xi}}{ 4 \pi^2 a^4 }   \int d  k \, k^3   e^{- 4 \sqrt{-2 \xi k \tau}} \left(1 - \frac{1}{4 \sqrt{-2 \xi k \tau}} \right)
\end{align}

\noindent Notice the integral is the same as the one above for $T_3$ and so

\begin{align}\label{eqeb}
\langle E \cdot B \rangle = - 2.09 \times 10^{-4} \, \frac{H^4 \, e^{2 \pi \xi}}{\xi^4 }  
\end{align}

\noindent This also means it cancels $T_3$ in the calculation of $\sigma^2$, which gives

\begin{align}\label{eqsigma}
\sigma^2_{E \cdot B} =  3.03 \times 10^{-8} \, \frac{H^8 \, e^{4 \pi \xi} }{\xi^8 }
\end{align}

\section{Finding $H$}

We can use the first Friedmann equation to write an expression for the Hubble parameter, $H^2 = \frac{\rho}{3 M_P^2}$. The kinetic energy of $\phi$  is $\frac{1}{2} \dot{\phi}^2$ when you ignore the spatial gradients, and the kinetic energy of the vector field is $\frac{1}{2} \langle E^2 + B^2 \rangle$.

\begin{align}
H^2 = \frac{1}{3 M_P^2} \left( \frac{1}{2} \dot{\phi}^2 + \frac{1}{2} \langle E^2 + B^2 \rangle + V \right)
\end{align}

First I'll find $\langle E^2(x) \rangle$ using that $E_i = - \frac{1}{a^2} A^{'}_i$ and using equations \ref{eqy} and \ref{eqz}.

\begin{align}
\langle A^{'}_i(k) \, A^{'}_i(k') \rangle =  u'(k)\,    u^{'*} (k) \, \delta^{(3)} (k + k')
\end{align}

\noindent giving

\begin{align}
\langle E^2(x) \rangle = \frac{1}{a^4 (2 \pi)^3} \int d^3 k   \, u'(k)\,    u^{'*} (k) 
\end{align}

\noindent Then we plug in for the mode functions, eq \ref{eqx}. 

\begin{align}
\langle E^2(x) \rangle = \frac{1}{a^4 2 \pi^2} \int_0^{\infty} d k \, k^2   \, \left( \frac{k \xi}{- 2 \tau} \right)^{\frac{1}{2}} e^{2 \pi \xi - 4 \sqrt{-2 \xi k \tau}} \left(1 - \frac{1}{4 \sqrt{-2  \xi k \tau}} \right)^2   
\end{align}

\noindent We can solve the integral and use that $a = e^{-N}$  and $\tau \approx - \frac{1}{H} e^{N}$: 

\begin{align}
 \langle E^2(x) \rangle = \frac{.0054 }{2 \pi^2 \sqrt{2}} \frac{H^4 e^{2 \pi \xi}}{ \xi^3 } 
\end{align}

Next we can solve for $\langle B^2(x) \rangle$ using that $B_i = \frac{1}{a^2} \epsilon_{ijk} \partial_j A_k$ and again using equations \ref{eqy} and \ref{eqz}.

\begin{align}
\langle B^2(x) \rangle = \frac{1}{a^4 (2 \pi)^3} \int d^3 k \int d^3 k' \, \epsilon_{ijk} \epsilon_{lmn}  \langle  (\partial_i e^{ i k \cdot x} A_j(k)) \cdot ( \partial_l e^{ i k' \cdot x} A_m(k') ) \rangle 
\end{align}

\begin{align}
\langle B^2(x) \rangle = \frac{1}{a^4 (2 \pi)^3} \int d^3 k \, k^2 \,  |u|^2
\end{align}

Then plug in for the mode functions, equation \ref{eqx}.

\begin{align}
\langle B^2(x) \rangle = \frac{1}{a^4 2 \pi^2 \cdot 2^{3/2}} e^{2 \pi \xi}  \left(\frac{- \tau}{\xi } \right)^{\frac{1}{2}} \int d^3 k \, k^{7/2} \,   e^{ - 4 \sqrt{-2 \xi k \tau}}
\end{align}

\begin{align}
\langle B^2(x) \rangle = \frac{.014 \, H^4}{ 2 \pi^2 \cdot 2^{3/2} \xi^5 } e^{2 \pi \xi} 
\end{align}

Putting both terms together gives:

\begin{align}
\langle E^2 + B^2 \rangle = \frac{H^4 \, e^{2 \pi \xi}}{2 \pi^2 \sqrt{2}} \left(\frac{.0054}{\xi^3} + \frac{.014}{2 \xi^5} \right)
\end{align}

Let:

\begin{align}
c =  \frac{ e^{2 \pi \xi}}{2 \pi^2 \sqrt{2}} \left(\frac{.0054}{\xi^3} + \frac{.014}{2 \xi^5} \right)
\end{align}

so 

\begin{align}
\langle E^2 + B^2 \rangle =c \, H^4
\end{align}

This allows us to write the Hubble parameter as:

\begin{align}\label{eqv}
H = \sqrt{ \frac{1}{2c} \left( 6 M_P^2 -  \phi_N^2 - \sqrt{( 6 M_P^2 -  \phi_N^2)^2 - 8 c V(\phi)} \right) }
\end{align}

\bibliographystyle{JHEPmodplain}
\bibliography{references}

\end{document}